\documentclass[onecolumn,11pt,aps,prd,nofootinbib,showpacs,preprintnumbers]{revtex4}
\usepackage{graphics}
\usepackage{multirow}
\usepackage{epsfig}
\usepackage{bbm}
\usepackage{bm}
\usepackage{amsfonts} 
\def\beq{\begin{equation}}
\def\eeq{\end{equation}}
\def\barr{\begin{eqnarray}}
\def\earr{\end{eqnarray}}

\def\lsim{\raise0.3ex\hbox{$\;<$\kern-0.75em\raise-1.1ex\hbox{$\sim\;$}}}
\def\gsim{\raise0.3ex\hbox{$\;>$\kern-0.75em\raise-1.1ex\hbox{$\sim\;$}}}

\def\rvec{{\bf r}}
\def\pvec{{\bf p}}

\def\Pvec{{\bf P}}
\def\Bvec{{\bf B}}
\def\Lvec{{\bf L}}

\def\Xvec{{\bf X}}
\def\Dvec{{\bf D}}
\def\Svec{{\bf S}}

\def\dmsq{\Delta m^2}
\def\msqa{\Delta m^2_{\rm atm}}
\def\msqs{\Delta m^2_{\odot}}

\def\nue{{\nu_e}}
\def\nuebar{{\bar{\nu}_e}}

\def\nux{{\nu_x}}
\def\nuxbar{{\bar{\nu}_x}}
\def\nuy{{\nu_y}}
\def\nuybar{{\bar{\nu}_y}}

\def\nubar{{\bar{\nu}}}
\def\la{\langle}
\def\ra{\rangle}

\begin{document}

\title{Collective three-flavor oscillations of supernova neutrinos}
\author{Basudeb Dasgupta and Amol Dighe}
\affiliation{Tata Institute of Fundamental Research,
Homi Bhabha Road, Mumbai 400005, India}

\begin{abstract}
Neutrinos and antineutrinos emitted from a core
collapse supernova interact among themselves,
giving rise to collective flavor conversion effects
that are significant near the neutrinosphere.
We develop a formalism to analyze these collective
effects in the complete three-flavor framework.
It naturally generalizes the spin-precession analogy
to three flavors and is capable of analytically
describing phenomena like vacuum/MSW oscillations,
synchronized oscillations, bipolar oscillations
and spectral split.
Using the formalism, we demonstrate that the flavor conversions
may be ``factorized'' into two-flavor oscillations
with hierarchical frequencies.
We explicitly show how the three-flavor solution
may be constructed by combining two-flavor
solutions.
For a typical supernova density profile, we identify 
an approximate separation of regions where distinctly 
different flavor conversion mechanisms operate, and
demonstrate the interplay between collective and
MSW effects.
We pictorialize our results in terms of the
``${\bf e}_3$--${\bf e}_8$ triangle'' diagram, which is a tool that
can be used to visualize three-neutrino flavor
conversions in general, and offers insights into
the analysis of the collective effects in
particular. 

\end{abstract}


\preprint{TIFR/TH/07-36}


\maketitle

\section{Introduction}                                   
\label{intro}

Neutrinos emitted from a core collapse supernova carry information 
about the primary fluxes, 
neutrino masses and mixing, and SN dynamics 
\cite{raffelt-0701677,amol-nufact}.
Neutrinos, produced in the region of the neutrinosphere, freestream outwards 
and pass through the core, mantle and envelope of the star. 
The drastically different environments in these regions,
consisting of varying densities of ordinary matter, neutrinos
and antineutrinos, affect flavor conversions among neutrinos.
Multiple shock fronts and turbulence generated during the 
explosion may also affect the neutrino flavor conversions.

The traditional picture of flavor conversions in a SN is based on 
the assumption that the effect of neutrino-neutrino interactions
is small. 
In this picture, neutrinos that are produced approximately as 
mass eigenstates at very high ambient matter density in the core 
propagate outwards from the neutrinosphere. 
Flavor conversion proceeds most efficiently at the electron 
density corresponding to the MSW resonance 
\cite{wolfenstein,mikheyev-smirnov}. 
The outcoming incoherent mixture of vacuum mass eigenstates is 
observed at a detector to be a combination of primary fluxes 
of the three neutrino flavors. 
This scenario of resonant neutrino conversions in a SN 
\cite{fuller-mayle-wilson-schramm-ApJ322} 
has been studied extensively to probe neutrino mixings and SN 
dynamics. 
The work has focussed on the determination of mass hierarchy and 
signatures of a non-zero $\theta_{13}$ 
\cite{dighe-smirnov-9907423,lunardini-smirnov-0302033}, 
earth matter effects on the neutrino fluxes when they
pass through matter \cite{dighe-keil-raffelt-jcap0306005,dighe-keil-raffelt-jcap0306006,dighe-kachelriess-raffelt-tomas-jcap0401},
shock wave effects on observable neutrino spectra and their
model independent signatures \cite{schirato-fuller-0205390,takahashi-sato-dalhed-wilson-0212195,fogli-lisi-mirizzi-montanino-0304056,tomas-kachelreiss-raffelt-dighe-janka-scheck-0407132,fogli-lisi-mirizzi-montanino-0412046,huber}. 
Recently, possible interference effects for multiple resonances 
\cite{dasgupta-dighe-0510219}, the role of turbulence in 
washing out shock wave effects 
\cite{fogli-lisi-mirizzi-montanino-0603033,choubey-harries-ross-0605255,friedland-gruzinov-0607244}, 
and time variation of the signal 
\cite{kneller-mclaughlin-brockman-0705.3853} 
have also been explored.

However, neutrino and antineutrino densities near the 
neutrinosphere are extremely high ($10^{30-35}$ per cm$^3$), which
make the neutrino-neutrino interactions significant.
Such a dense gas of neutrinos and antineutrinos is coupled to 
itself, making its evolution nonlinear.
The flavor off-diagonal terms can be sizeable, 
and significant flavor conversion is possible 
\cite{pantaleone-PRD46,pantaleone-PLB287}. 
A formalism to study flavor evolution of such dense relativistic 
neutrino gases was developed in 
\cite{thompson-mckellar-PLB259,raffelt-sigl-NPB406,thompson-mckellar-PRD49}, 
where a set of quantum kinetic equations for their evolution 
were written down. 
These equations have been studied in detail, though mostly in 
the two-flavor approximation, and the nature of flavor evolution 
has been identified \cite{samuel-PRD48,kostecky-samuel-9506262,pantaleone-PRD58,samuel-9604341}. 
It was eventually understood that a dense gas of neutrinos 
displays collective flavor conversion, i.e. neutrinos of all 
energies oscillate together, through synchronized oscillations 
\cite{pastor-raffelt-semikoz-0109035} and/or bipolar oscillations 
\cite{hannestad-raffelt-sigl-wong-0608095,duan-fuller-carlson-qian-0703776}. 
Another remarkable effect of these interactions is a 
partial or complete swapping of the energy spectra of two 
neutrino flavors, called step-wise spectral swapping or 
simply spectral splits, as the neutrinos transit from a 
region where collective effects dominate to a region where 
neutrino density is low 
\cite{raffelt-smirnov-0705.1830,raffelt-smirnov-0709.4641}.

The nonlinear effects in the context of SNe were considered in 
\cite{pantaleone-9405008,qian-fuller-9406073,sigl-9410094,pastor-raffelt-0207281,balantekin-yuksel-0411159}. 
Recent two-flavor simulations showed that the collective
effects affect neutrino flavor conversions substantially
\cite{duan-fuller-carlson-qian-0606616,duan-fuller-carlson-qian-0608050}. 
Different collective flavor transformations seem to play 
a part in different regions of the star. 
Many features of the results of these simulations can be 
understood from the ``single-angle'' approximation, 
i.e. ignoring the dependence of the initial launching angle 
of neutrinos on the evolutions of neutrino trajectories 
\cite{fuller-qian-0505240,duan-fuller-qian-0511275,hannestad-raffelt-sigl-wong-0608095,duan-fuller-carlson-qian-0703776,raffelt-smirnov-0705.1830,duan-fuller-qian-0706.4293,duan-fuller-carlson-zhong-0707.0290,estebanpretel-pastor-tomas-raffelt-sigl-0706.2498,fogli-lisi-marrone-mirizzi-0707.1998,raffelt-smirnov-0709.4641}. 
Angular dependence of flavor evolution can give rise to 
additional angle dependent features observed in two-flavor 
simulations \cite{duan-fuller-carlson-qian-0606616,duan-fuller-carlson-qian-0608050}, 
or to decoherence effects 
\cite{pantaleone-PRD58,raffelt-sigl-0701182}. 
For a realistic asymmetry between neutrino and antineutrino 
fluxes, such angle dependent effects are likely to be small 
\cite{estebanpretel-pastor-tomas-raffelt-sigl-0706.2498,fogli-lisi-marrone-mirizzi-0707.1998}. 
Recently collective effects have also been numerically 
investigated in the context of the neutronization-burst phase 
of O-Ne-Mg supernovae \cite{duan-fuller-carlson-qian-0710.1271}. 
The impact of these nonlinear interactions has also been studied 
in the context of cosmological neutrino flavor equilibration in 
the early Universe where the synchronized oscillations play 
a significant part \cite{kostelecky-pantaleone-samuel-PLB315,kostelecky-pantaleone-samuel-PLB318,kostelecky-pantaleone-samuel-PRD49,kostelecky-samuel-9507427,kostelecky-samuel-9610399,dolgov-hansen-pastor-petkov-raffelt-semikoz-0201287,wong-0203180,wong-talk,abazajian-beacom-bell-0203442}.

Most of the analytical results in this area are in the 
two-flavor approximation, where the equations describing
flavor transformations are similar to the 
equations of motion of a gyroscope. 
As a result, two-flavor results are now fairly well understood 
analytically, with the exception of possibly two issues, 
{\it viz.} decoherence (or lack of it) for asymmetric systems 
\cite{estebanpretel-pastor-tomas-raffelt-sigl-0706.2498}, 
and the existence of the antineutrino spectral split 
\cite{fogli-lisi-marrone-mirizzi-0707.1998}. 
In this work we focus on the effects of the mixing of the 
third flavor. 
We present an analytical framework to study three-flavor effects 
and demonstrate an approximate factorization of the full 
three-flavor problem into simpler two-flavor problems, 
when $\msqs\ll|\msqa|$ and $\theta_{13} \ll 1$. 
We also develop the ``${\bf e}_3$--${\bf e}_8$ triangle''
diagram, a tool that can be used to visualize the three
neutrino flavor conversions in general, and allows one to
gauge the extent of additional effects of the third flavor.
We numerically study collective flavor transformations 
for a typical SN density profile, 
identify regions where different flavor conversion 
mechanisms operate, and explain the features of the spectra 
using our formalism.

The outline of this paper is as follows. 
In Sec.~\ref{formalism}, we review the equations of motion of 
a dense gas of neutrinos in steady state. 
We specialize these equations to spherical geometry in the 
single-angle half-isotropic approximation, and write the 
three-flavor analogue of the gyroscope equations,
by introducing the eight-dimensional Bloch vectors.
We recover the two-flavor limit of those equations, and recognize 
an approximate factorization of the three-flavor problem 
(reminiscent of the $H$-$L$ factorization in the standard picture)
into smaller two-flavor problems. 
We show that survival probabilities can be written down in 
a simple form, purely in terms of the solutions to the 
two-flavor problems, as long as the frequencies governing
the oscillations are hierarchically separated. 
In Sec.~\ref{toy-cases}, we illustrate the above factorization
for vacuum/MSW oscillations as well as collective synchronized
oscillations. 
We also explain the three-flavor features of bipolar oscillations
and spectral splits qualitatively and pictorially.
In Sec.~\ref{full-sn}, we calculate the flavor conversion
probabilities numerically for a typical SN matter profile, 
and identify the additional features due to the third neutrino.
We conclude in Sec.~\ref{concl} with a summary of our results 
and comments about directions of future investigation. 

\section{Three-Flavor Formalism}                                   
\label{formalism}

In this section we derive the steady state equations of motion for 
an ensemble of dense gas of three-flavor neutrinos, as a 
straightforward generalization of the corresponding equations in
 the two-flavor case.

\subsection{Hamiltonian and Equations of Motion}
\label{notation}

We denote a neutrino of momentum $\pvec$ at time $t$ at position 
$\rvec$ by $\nu(\pvec,\rvec,t)$,
and work in the modified flavor basis $(\nu_e, \nu_x, \nu_y)$
defined such that $(\nu_e ~\nu_x ~\nu_y)^T = 
R_{23}^\dagger (\theta_{23})
(\nu_e ~\nu_\mu ~\nu_\tau)^T$, where $R_{23}$ is the rotation
matrix and $\theta_{23}$ the neutrino mixing angle in the
$2$-$3$ plane. \footnote{This basis has also been denoted in the 
literature as $(\nu_e, \nu_{\mu'}, \nu_{\tau'})$ 
\cite{dighe-smirnov-9907423}.}
In this basis, the density matrix for $n_{\nu}(\pvec,\rvec,t)$ neutrinos 
with momenta between $\pvec$ and $\pvec+d\pvec$
at any position between  $\rvec$ and $\rvec+d\rvec$ 
may be written as
\barr
{\rho}_{\nu_{\alpha}\nu_{\beta}}(\pvec,\rvec,t) & \equiv& 
\frac{1}{n_{\nu}(\pvec,\rvec,t)}\sum|\nu(\pvec,\rvec,t) \ra 
\la \nu(\pvec,\rvec,t)|_{\alpha\beta} ,
\earr
where $\alpha,\beta = e,x,y$ and the summation is over all $n_{\nu}(\pvec,\rvec,t)$ neutrinos.
Note that the density matrix is normalized to have unit trace, 
but the neutrino density itself is $n_{\nu}(\pvec,\rvec,t)$,
which typically falls off as $1/r^2$ from the source. 
The number density of neutrinos with flavor $\alpha$ is obtained
through
\beq
n_{\nu_\alpha}(\pvec,\rvec,t)=
n_{\nu}(\pvec,\rvec,t)\rho_{\nu_\alpha \nu_\alpha}(\pvec,\rvec,t)~.
\label{nevsn}
\eeq
We define the analogous quantities for antineutrinos similarly, 
but with a reversed order of flavor indices to keep the form of 
equations of motion identical \cite{raffelt-sigl-NPB406}, 
and denote them with a ``bar'' over the corresponding variables 
for neutrinos.

The effective Hamiltonian in the modified flavor basis for neutrinos 
$\nu(\pvec,\rvec,t)$ of energy $E \approx p = |\pvec|$ in vacuum is 
\beq
H_{vac}(p)= UM^2U^{\dagger}/2p \; ,
\eeq 
where the masses and the mixing matrix are parametrized as
\barr
M& \equiv &{\rm Diag}(m_1,m_2, m_3)~,\label{masses}\\
U& \equiv &R_{23}^\dagger(\theta_{23}) R_{23}(\theta_{23})
R_{13}(\theta_{13})R_{12}(\theta_{12})~,
\label{mixingmatrix}
\earr 
with $R_{ij}$ being the rotation matrices in the $i$-$j$ plane.
In this work, we take the value of the $CP$-violating phase in neutrino sector to be zero.
Now $H_{vac}(p)$ may be explicitly written as 
\barr
H_{vac}(p)&=&\frac{\Delta m^2_{13}}{2p}\left(
\begin{array}{ccc}
s^2_{13} & 0 & c_{13}s_{13}\\
0 & 0 & 0 \\
c_{13} s_{13} & 0 & c^2_{13}
 \end{array}\right)\nonumber
\\
&+&\frac{\Delta m^2_{12}}{2p}\left(
\begin{array}{ccc}
 c^2_{13}s^2_{12} & c_{12}c_{13}s_{12} & -c_{13}s^2_{12}s_{13}\\
 
c_{12}c_{13}s_{12} & c^2_{12} & -c_{12}s_{12}s_{13} \\

-c_{13}s^2_{12}s_{13} & - c_{12}s_{12}s_{13} & s^2_{12}s^2_{13}
 \end{array}
\right)~,
\earr
where $\dmsq_{ij}=m_j^2-m_i^2$ and 
other symbols have their usual meanings. 
In matter, neutrinos feel the Mikheyev-Smirnov-Wolfenstein (MSW) 
potential \cite{wolfenstein,mikheyev-smirnov} due to charged leptons \footnote{We assume that the density of $\mu^\pm$ and $\tau^\pm$ 
is negligible, and that $\nu_\mu$ and $\nu_\tau$ feel approximately
identical potentials, which have been taken to be zero by
convention.
An analysis of collective effects including a $\mu-\tau$ potential
has recently been carried out \cite{EstebanPretel:2007yq}.} 
\beq
V(\rvec,t)=\sqrt{2}G_F\;n_{e^-}(\rvec,t)\;\rm{Diag}(1,0,0)
\label{msw-potential}
\eeq
that adds to the Hamiltonian, 
where $n_{e^{-}}(\rvec)$ is the net electron number density at $\rvec$.
The effective Hamiltonian also includes the effects of 
neutrino-neutrino interactions, which to the leading order 
in $G_F$ depend only on forward scattering and contribute
\cite{thompson-mckellar-PLB259,raffelt-sigl-NPB406,thompson-mckellar-PRD49}
\beq
H_{\nu \nu}(\pvec,\rvec,t)=
\sqrt{2}G_F\int \frac{d^3{\bf q}}{\left(2\pi\right)^3}
\kappa_{\pvec {\bf q}}
\bigg(n_{\nu}({\bf q},\rvec,t)\rho({\bf q},\rvec,t)-
\bar{n}_{\nu}({\bf q},\rvec,t)\bar{\rho}({\bf q},\rvec,t)\bigg)~.
\label{Hnunu}
\eeq
The interaction strength is dependent on the angular separation of 
the momenta of the interacting particles, and is given by 
$\kappa_{\pvec {\bf q}} \equiv 1-\cos{\theta_{\pvec {\bf q}}}$, 
where $\theta_{\pvec {\bf q}}$ is the angle between 
$\pvec$ and ${\bf q}$.

The equation of motion for the density matrix is
\barr
\frac{d}{dt}\rho(\pvec,\rvec,t)
&=& -i \bigg[ H(\pvec,\rvec,t),\rho(\pvec,\rvec,t) \bigg] + 
\frac{\partial}{\partial t}\rho(\pvec,\rvec,t) \; .
\earr
In the steady state (no explicit time dependence in the Hamiltonian 
and initial conditions) and ignoring external forces 
(terms depending on $d\pvec/dt$), 
we can drop the time dependence in the problem. 
Writing ${\bf {v}}=d\rvec/dt$  we have the equations of motion 
for $\rho(\pvec,\rvec)$ and $\bar{\rho}(\pvec,\rvec)$ as \cite{Cardall:2007zw} 
\barr
{\bf {v}} \cdot \partial_{\rvec} \rho(\pvec,\rvec)
&=&-i \bigg[ +H_{vac}(p)+V(\rvec)+H_{\nu\nu}(\pvec,\rvec),
\rho(\pvec,\rvec) \bigg]~,\label{gen-eom1}\\
{\bf {v}} \cdot \partial_{\rvec} \bar{\rho}(\pvec,\rvec)
&=&-i \bigg[-H_{vac}(p)+V(\rvec)+H_{\nu\nu}(\pvec,\rvec),
\bar{\rho}(\pvec,\rvec) \bigg]~.
\label{gen-eom2}
\earr

The effect of ordinary matter can be ``rotated away'' 
by working in the interaction picture 
\cite{duan-fuller-qian-0511275,duan-fuller-carlson-qian-0606616}. 
We employ an operator ${O(\rvec)}$ under which a matrix $A$ 
transforms to 
\beq
A^{int}(\rvec)={O(\rvec)}A {O}^{-1}(\rvec)~,
\label{simtransf}
\eeq  
where
\beq
{O(\rvec)}=\exp\left( {i \int_{0}^{\rvec} d\rvec' V(\rvec')}\right)~.
\label{O-def}
\eeq
This choice simplifies the equations of motion 
by removing the matter term, giving us
\barr
{\rm{\bf v}} \cdot \partial_{\rvec} \rho^{int}(\pvec,\rvec)
&=&-i \bigg[+H_{vac}^{int}(p,\rvec)+H_{\nu\nu}^{int}(\pvec,\rvec),
\rho^{int}(\pvec,\rvec) \bigg]~,\label{gen-eom1v}\\
{\rm{\bf v}} \cdot \partial_{\rvec} \bar{\rho}^{int}(\pvec,\rvec)
&=&-i \bigg[-H_{vac}^{int}(p,\rvec)+H_{\nu\nu}^{int}(\pvec,\rvec),
\bar{\rho}^{int}(\pvec,\rvec) \bigg]~.
\label{gen-eom2v}
\earr
The transformation by ${O(\rvec)}$ leaves diagonal entries of 
$\rho(\pvec,\rvec),\;\bar{\rho}(\pvec,\rvec), H_{vac}(p)$ 
and $H_{\nu \nu}(\pvec, \rvec)$ unchanged, but the off-diagonal 
entries become $\rvec$-dependent. 
For example, if $V(\rvec)$ varies adiabatically and only 
in the radial direction, the vacuum Hamiltonian changes 
according to Eq. (\ref{simtransf}) as
\beq
H_{vac}^{int}(p,r) = H_{vac}(p) + ir \bigg[V(r),H_{vac}(p) \bigg]+
\frac{(ir)^2}{2!} \bigg[V(r), \bigg[V(r),H_{vac}(p) \bigg] \bigg]+...~.
\label{hoint}
\eeq
We know that $V(r)$ is a diagonal matrix, so 
only the off-diagonal elements of $H_0(p)$ are affected by the 
transformation. 
The final observables we are going to be interested in, the
number fluxes of neutrino flavors, involve only diagonal
elements of the density matrix [see Eq.~(\ref{nevsn})],
so the interaction basis is well suited for our purposes.

\subsection{Spherical Symmetry and Single-angle Equations of Motion}
\label{single-angle}

The interaction term $H_{\nu\nu}$ in Eq. (\ref{Hnunu}) 
depends on $\theta_{\pvec {\bf q}}$, i.e. the angle between 
the momenta of interacting neutrinos. 
Thus while performing the angular integrals therein, 
the dependence of the neutrino flux on all angular variables 
must be taken into account. This makes the problem quite complicated, 
and an approximate treatment is needed in order to gain  
useful insights. 
Two levels of approximation have been considered in literature, 
{\it viz.} multi-angle and single-angle. 
In the multi-angle approximation, azimuthal symmetry about 
the axis defined by the source and observer is usually assumed, 
but not complete spherical symmetry. 
This essentially captures the effects of correlations between 
trajectories with different initial launching angles. 
The effects of such correlations can have interesting consequences 
which have been explored in detail 
\cite{duan-fuller-carlson-qian-0606616,duan-fuller-carlson-qian-0608050,pantaleone-PRD58,raffelt-sigl-0701182,estebanpretel-pastor-tomas-raffelt-sigl-0706.2498,fogli-lisi-marrone-mirizzi-0707.1998}. 
In the single-angle approximation, it is assumed that the flavor 
evolution does not significantly depend on any of the angular 
coordinates (i.e. the evolution is spherically symmetric), 
and thus we can integrate over all angular degrees of freedom 
trivially. 
One must then choose a representative value for 
$\cos \theta_{\pvec}$, which we take to be $1/2$.

We assume half-isotropic emission from a source of radius $r_{0}$, 
as defined in 
\cite{estebanpretel-pastor-tomas-raffelt-sigl-0706.2498}, and write 
\barr
n_{\nu}(\pvec,\rvec)&=& n_\nu(p,r) =
n_{\nu}(p,r_{0}) ~ r_{0}^2/r^2~,\\
\rho(\pvec,\rvec)&=&\rho(p,r)~.
\earr 
In the steady state, the fluxes of neutrinos and antineutrinos 
can be written as
\barr
\Phi_{\nu}&=&\int dp\; 2\pi p^2\; 4 \pi r_0^2 \; n_{\nu}(p,r_{0})~,\\
\Phi_{\nubar}&=&\int dp\; 2\pi p^2\; 4 \pi r_0^2 \; 
\bar{n}_{\nu}(p,r_{0}) ~,
\earr
the total flux being $\Phi = \Phi_\nu + \Phi_\nubar$.

A further ``unification'' in the notation for neutrinos
and antineutrinos is possible by noting that their equations 
of motion, i.e. Eqs.~(\ref{gen-eom1}) and (\ref{gen-eom2}), 
differ only in the sign of $H_{vac}(p)$. 
This suggests a change of variables from $p$ to
\beq
\omega=|\dmsq_{13}|/(2p) \; .
\label{omegadefn}
\eeq 
Using the same convention as \cite{raffelt-smirnov-0705.1830},
we define for neutrinos
\beq
n_\nu(\omega,r) \equiv n_\nu (~p(\omega),~r ) \; , \quad
\rho(\omega,r) \equiv \rho( ~p(\omega),~r) \; , 
\label{new-nnu}
\eeq
and for antineutrinos
\beq
n_\nu(-\omega,r) \equiv \bar{n}_\nu(~p(\omega),~r) \; , \quad
\rho(-\omega,r) \equiv \bar{\rho}(~p(\omega),~r) \; . 
\label{new-nnubar}
\eeq
The negative values of $\omega$ thus correspond to antineutrinos.
Then we need to solve only for $\rho(\omega,r)$, albeit at the cost 
of extending the domain of $\omega$ to both positive and negative 
values.
This simplifies the $H_{\nu\nu}(\pvec,\rvec)$ term in 
Eq. (\ref{Hnunu}) to \footnote{
Note that $H_{\nu \nu}(\pvec,\rvec)$ depended on $\pvec$ only
through the direction of $\pvec$. This dependence no longer
survives in the single-angle approximation.}
\beq
H_{\nu\nu}(r)=\mu(r)\int_{-\infty}^{\infty} d\omega\;f({\omega})\;\rho({\omega},r)\;sgn(\omega)~.
\label{hnunu-r}
\eeq
in terms of the distribution function
\beq
f(\omega)= \frac{1}{\Phi}
\frac{|\dmsq_{13}|^3 \pi^2 r_0^2}{\omega^4} \;
n_{\nu}(\omega,r_{0})~,
\label{fomega}
\eeq
normalized as $\int_{-\infty}^{\infty} d\omega\;f(\omega)=1$,
and the ``collective potential''
\beq
\mu(r) = \mu_0 ~g(r) \; .
\eeq
Here $\mu_0$ is the collective potential at the neutrinosphere:
\beq
\mu_0 \equiv \mu(r_0) = \frac{3\sqrt{2}G_F\Phi}{128 \pi^4 r_0^2}\; ,
\label{mu0}
\eeq
and the ``geometric dilution factor'' $g(r)$ is given by
\barr
g(r) & \equiv & \frac{4 r_0^2}{3 r^2}
\int^{1}_{\sqrt{1 - (r_0/r)^2}} 
d (\cos \theta_{{\bf q}})\;(1-\cos \theta_{{\bf q}} 
\cos \theta_{\pvec}){\bigg \vert}_{\cos\theta_{\pvec}=1/2} \; \nonumber \\
& = & \frac{4 r_0^2}{3 r^2}\left(1 - \sqrt{1 - \frac{r_0^2}{r^2}} - 
\frac{r_0^2 }{4 r^2}\right)
 \; .
\label{dilfac}
\earr
The geometric dilution factor equals unity for $r=r_0$,
whereas at large $r_0$, it decreases as $1/r^4$.
The decrease of neutrino densities from a finite source
accounts for a factor of $1/r^2$, whereas
the additional dilution factor of approximately $1/r^2$ 
arises from the integral in Eq.~(\ref{dilfac}), 
due to the decreasing angle subtended by the source and 
reduced collinearity, which are encoded in the limits and the 
integrand respectively
\cite{fuller-qian-0505240}. 
Note that the exact numerical factors depend on
the choice of $\cos \theta_{\pvec}$.

The total flux $\Phi$ remains conserved as long as there is no 
explicit temporal variation of the overall luminosity. 
We work in the steady state approximation and assume the 
luminosity to be constant in time. 
Slow variations in it may be taken into account by including 
an additional time dependent factor. 
Note that $f(\omega)$ is independent of $r$, 
which embodies the fact that the normalized neutrino spectrum 
does not change. 
Using Eq. (\ref{nevsn}), we can also write the flavor dependent 
$\omega$-spectra $f_{\nu_{\alpha}}(\omega,r)$ as
\beq
f_{\nu_\alpha}(\omega,r)=f(\omega)\rho_{\nu_\alpha \nu_\alpha}
(\omega,r)~.
\eeq
Note that $f_{\nue}(\omega,r)$ contains the spectra of both $\nue$ and $\nuebar$, and depends on $r$ only through $\rho_{\nue\nue}(\omega,r)$. It would be a constant on the trajectory if there were no flavor evolution of $\rho_{\nue\nue}(\omega,r)$. For later use, we define the energy integrated neutrino fluxes for each flavor as
\barr
\Phi_{\nue}(r)&=&\Phi\int_{0}^{\infty} d\omega\;f_{\nue}(\omega,r)~,\\
\Phi_{\nu}&=&\Phi_{\nue}(r) + \Phi_{\nu_x}(r) + \Phi_{\nu_y}(r) ~,\\
\Phi_{\nuebar}(r)&=&\Phi\int_{-\infty}^{0} d\omega\;
f_{\nue}(\omega,r)~,\\
\Phi_{\nubar}&=&\Phi_{\nuebar}(r) + \Phi_{\bar{\nu}_x}(r) + 
\Phi_{\bar{\nu}_y}(r)~.
\earr

With these approximations, the problem is reduced to an 
ordinary one dimensional problem along the radial direction. 
We denote the derivative with respect to $r$ 
using a ``dot'', and using Eqs.~(\ref{gen-eom1}) and (\ref{gen-eom2}), 
arrive at the single-angle equations of motion
\beq
v_r \dot{\rho}(\omega,r)=-i \bigg[ +H_{vac}(\omega,h)
+V(r)+H_{\nu\nu}(r),\rho(\omega,r) \bigg]~,
\label{sageom-old}
\eeq
where $h \equiv sgn(\dmsq_{13})$ encodes the dependence on mass 
hierarchy. 
Here 
\beq
v_r = \sqrt{ 1 - \frac{r_0^2}{r^2} \sin^2 \theta_{\pvec}(r_0)}
\eeq
is the radial velocity of the neutrino.
Note that for $r \gg r_0$, we have $v_r \approx 1$.
Since the flavor conversions due to collective effects start
becoming significant only for $r > 4 r_0$ 
\cite{estebanpretel-pastor-tomas-raffelt-sigl-0706.2498,fogli-lisi-marrone-mirizzi-0707.1998},
for our analytic approximations we take $v_r = 1$,
making Eq. (\ref{sageom-old})
\beq
\dot{\rho}(\omega,r)=-i \bigg[ +H_{vac}(\omega,h)
+V(r)+H_{\nu\nu}(r),\rho(\omega,r) \bigg]~.
\label{sageom}
\eeq

We have thus used the spherical symmetry of the problem, 
and the simple energy dependence, to rephrase the equations 
of motion in a somewhat simpler form. 
This single-angle approximation is probably crude, 
but it has been shown in numerical simulations 
(for two flavors) that this approximation seems to work reasonably 
well \cite{fogli-lisi-marrone-mirizzi-0707.1998}. 
It also seems that the multi-angle effects are suppressed 
when the neutrino and antineutrino spectra are not identical 
\cite{estebanpretel-pastor-tomas-raffelt-sigl-0706.2498}.
We assume the above results to hold true for three flavors as well,
and ignore multi-angle effects in this work. 
Thus, for an analytical understanding of various flavor 
conversion phenomena associated with this system, 
we confine our discussion to the steady-state single-angle 
half-isotropic approximation that we have outlined above. 

\subsection{Bloch Vector Notation}
\label{bloch}

In the single-angle approximation, it is useful to 
re-express the density matrices and the Hamiltonian as 
Bloch vectors. 
The idea, analogous to the two-flavor case, is to express the 
matrices in a basis of hermitian matrices, 
and to study the motion of the vectors constructed out of 
the expansion coefficients (which are called the Bloch vectors). 
In our problem, we choose the basis consisting of 
the 3$\times$3 identity matrix $I$, and the $8$ Gell-Mann 
matrices $\Lambda_{a}$ given by
\barr
\Lambda_1&=&\left[\begin{array}{ccc}
0 & 1 & 0\\
1 & 0 & 0\\
0 & 0 & 0\\
\end{array}\right],~ 
\Lambda_2=\left[\begin{array}{ccc}
0 & -i & 0\\
i & 0 & 0\\
0 & 0 & 0\\
\end{array}\right],~ 
\Lambda_3=\left[\begin{array}{ccc}
1 & 0 & 0\\
0 & -1 & 0\\
0 & 0 & 0\\
\end{array}\right],~\nonumber\\
\Lambda_4&=&\left[\begin{array}{ccc}
0 & 0 & 1\\
0 & 0 & 0\\
1 & 0 & 0\\
\end{array}\right],~
\Lambda_5=\left[\begin{array}{ccc}
0 & 0 & -i\\
0 & 0 & 0\\
i & 0 & 0\\
\end{array}\right],~ 
\Lambda_6=\left[\begin{array}{ccc}
0 & 0 & 0\\
0 & 0 & 1\\
0 & 1 & 0\\
\end{array}\right],~\nonumber\\
\Lambda_7&=&\left[\begin{array}{ccc}
0 & 0 & 0\\
0 & 0 & -i\\
0 & i & 0\\
\end{array}\right],~
\Lambda_8=\frac{1}{\sqrt{3}}\left[\begin{array}{ccc}
1 & 0 & 0\\
0 & 1 & 0\\
0 & 0 & -2\\
\end{array}\right],~ 
\earr
which satisfy the $SU(3)$ Lie algebra 
\beq
[\Lambda_a,\Lambda_b]=if_{abc}\;\Lambda_c~,
\eeq
where $a,b,c$ are integers from 1 to 8.
Note that the normalization for the matrices is chosen such that
\beq
{\rm Tr}(\Lambda_{a}\Lambda_{b})=2\delta_{ab}~.
\label{trace}
\eeq
The structure constants $f_{abc}$ are antisymmetric under 
exchange of any two indices and are specified by 
\beq
f_{123}=2 \; ; \quad 
f_{147},f_{165},f_{246},f_{257},f_{345},f_{376}=1 \; ; \quad
f_{678},f_{458}=\sqrt{3} \; .
\label{fijk}
\eeq  
Note that basis of traceless matrices $\Lambda_a$ 
can be expressed as a semi-direct sum of 
\beq
{\mathbbm K}=\{\Lambda_1,\Lambda_2,\Lambda_3,\Lambda_8\}\; \qquad
{\rm and}\; \qquad 
{\mathbbm Q}=\{\Lambda_4,\Lambda_5,\Lambda_6,\Lambda_7\}~,
\eeq
i.e. for $K_{a}\in {\mathbbm K}$ and $Q_{a}\in {\mathbbm Q}$ 
we have
\beq
[K_a,Q_b]\in  {\mathbbm K}~,~ [Q_a,Q_b]\in {\mathbbm K}  
\quad {~\rm and~} \quad [Q_a,K_b] \in {\mathbbm Q}~.
\label{kqdef}
\eeq
In fact this is not the only choice of ${\mathbbm K}$ 
and ${\mathbbm Q}$ that has this property. 
In addition to the decomposition
\beq
{\mathbbm K}^{ex}=\{\Lambda_1,\Lambda_2,\Lambda_3,\Lambda_8\}\; \qquad
{\rm and}\; \qquad 
{\mathbbm Q}^{ex}=\{\Lambda_4,\Lambda_5,\Lambda_6,\Lambda_7\}~,
\label{Kex}
\eeq
as above, we could also choose  
\barr
{\mathbbm K}^{ey} &=&
\{\Lambda_3,\Lambda_4,\Lambda_5,\Lambda_8\}\; 
\quad {\rm and}\; \quad 
{\mathbbm Q}^{ey} =\{\Lambda_1,\Lambda_2,\Lambda_6,\Lambda_7\}\; 
\qquad {\rm or}\;
\label{Key} \\  
{\mathbbm K}^{xy}&=&\{\Lambda_3,\Lambda_6,\Lambda_7,\Lambda_8\}\; 
\quad {\rm and}\; \quad
{\mathbbm  Q}^{xy}=\{\Lambda_1,\Lambda_2,\Lambda_4,\Lambda_5\}~,
\label{Kxy}
\earr
which satisfy the conditions in Eq. (\ref{kqdef}). 
The meaning of the superscripts ($ex,ey,xy$) 
on ${\mathbbm K}$ and ${\mathbbm Q}$ will become clear later.

\begin{figure}
\centering
\includegraphics[width=15.0cm]{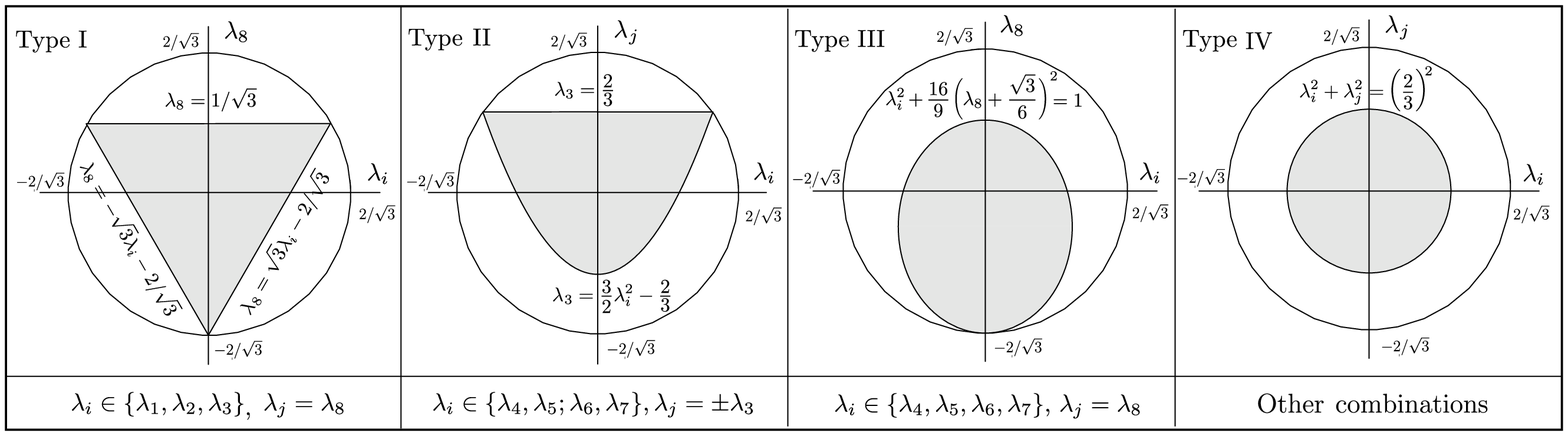}
\caption{The shape of the Bloch-ball for a vector 
$\lambda_i \hat{\bf e}_i$ (Figure taken 
from \cite{blochball-figure}).}
\label{bloch-ball}
\end{figure}

Using the basis matrices $I$ and $\Lambda_a$, 
we now express any $3 \times 3$ hermitian matrix $X$ as 
a vector $\Xvec$ in the $SU(3)$ generator space 
(with unit vectors ${\hat{\bf e}_i}$) as
\beq
X=\frac{1}{3}\;{\rm X}_0\;I +\frac{1}{2}\;\Xvec \cdot {\bm \Lambda}~.
\label{reparam}
\eeq
We call $\Xvec$ the Bloch vector corresponding to the matrix $X$.
The vector $\Xvec$ must lie completely within an
eight-dimensional compact volume, called the Bloch ball, 
whose various two-dimensional sections are shown in 
Fig.~\ref{bloch-ball}. 
We say that the vector $\Xvec$ is contained in 
${\mathbbm K}^{ex}$ 
(${\mathbbm K}^{ey}$, ${\mathbbm K}^{xy})$ 
if the matrix $X$ can be expressed solely as a linear 
combination 
of $\Lambda_{a}\in{\mathbbm K}^{ex}$ 
(${\mathbbm K}^{ey}$, ${\mathbbm K}^{xy})$.

We reparametrize our equations using Eq. (\ref{reparam}), 
and define the Bloch vectors corresponding to the density matrices as  
\beq
\rho(\omega,r)=\frac{1}{3}\;{\rm P}_0(\omega,r)\;I +
\frac{1}{2}\;\Pvec(\omega,r) \cdot  {\bm \Lambda}~.
\label{polvec} 
\eeq
Note that $\bm \Lambda$ is an eight-vector of $3\times 3$
matrices. 
The scalar ${\rm P}_0(\omega,r)$ and the polarization vector 
$\Pvec(\omega,r)$ encode the flavor content of neutrinos of 
energy $p=|\dmsq_{13}|/(2\omega)$ at a distance $r$ for $\omega>0$. 
The negative values of $\omega$ encode the same information 
for antineutrinos. 
Since $\rho(\omega,r)$ has been normalised to have unit trace by definition, 
${\rm P}_0(\omega,r)$ is equal to one. 
We will therefore not worry about the zeroth component of the 
polarization vector henceforth.
For a pure state, $\Pvec(\omega,r)$ lies on the boundary of
the shaded region in Fig.~\ref{bloch-ball}, and has the
magnitude $2/\sqrt{3}$.
For a mixed state, the magnitude of  $\Pvec(\omega,r)$ is
smaller and the vector lies within the shaded region.

We assume that all neutrinos are produced as flavor eigenstates, 
i.e. the primary flux consists of 
$n_{\nu_{\alpha}}(p,r_{0})$ and $\bar{n}_{\nu_{\alpha}}(p,r_{0})$ 
with energy $p$. 
The initial density matrix $\rho(p,r_{0})$ is therefore 
${\rm Diag}\bigg(n_{\nue}(p,r_{0}),n_{\nu_x}(p,r_{0}),
n_{\nu_y}(p,r_{0}) \bigg)$,
and similarly for antineutrinos. 
The initial polarization vector may be written as
\beq
\Pvec(\omega,r_{0})=\frac{f_{\nue}(\omega,r_{0})
-f_{\nu_x}(\omega,r_{0})}{f(\omega)}\;\hat{\bf e}_3 + 
\frac{f_{\nu_e}(\omega) + f_{\nu_x}(\omega)- 
2f_{\nu_y}(\omega,r_{0})}{\sqrt{3}\;f(\omega)}\;\hat{\bf e}_8\;.
\label{pvecinitial}
\eeq

\begin{figure}
\epsfig{file=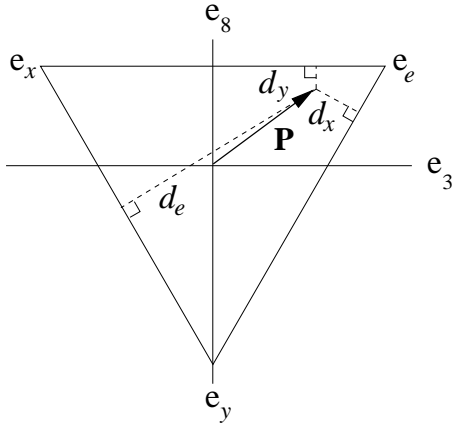,width=6.0cm}
\caption{The projection of a polarization vector 
$\Pvec$ on the ${\bf e}_{3}$--${\bf e}_{8}$ plane}
\label{tria}
\end{figure}

The polarization vector $\Pvec(\omega,r)$, 
when projected onto the ${\bf e}_{3}$--${\bf e}_{8}$ plane,
must lie within the triangle in Fig.~\ref{tria}, where
we show a representative $\Pvec(\omega,r)$ projected on 
the ${\bf e}_{3}$--${\bf e}_{8}$ plane.
The pure electron flavor is represented by
\beq
{\bf e}_e = \hat{\bf e}_3 + \frac{\hat{\bf e}_8}{\sqrt{3}} \; .
\eeq  
The $\nue$ or $\nuebar$ content with energy $p$ at 
position $r$ is given by
\beq
\rho_{\nue\nue}(p,r)=\frac{n_{\nue}(p,r)}{n_\nu(p)}
=\frac{f_{\nue}(\omega,r)}{f(\omega)}
=\frac{1}{3}+ \frac{\Pvec \cdot {\bf e}_{e}}{2}
=\frac{d_{e}}{\sqrt{3}}~.
\label{rhoeefinal}
\eeq
The projection of $\Pvec$ on $\hat{\bf e}_{e}$ is thus related to 
$\rho_{\nue\nue}(p,r)=f_{\nue}(\omega,r)/f(\omega)$ as above.
The same quantity can be easily visualized from the figure as 
$d_{e}/\sqrt{3}$, where $d_e$ is the distance of the tip of 
$\Pvec$ from the side of the triangle that is perpendicular to 
$\hat{\bf e}_{e}$ (as shown in the figure). 
The number of $\nu_x$ and $\nu_y$ are also similarly calculated. 
Negative values of $\omega$ encode the same information for the 
antineutrinos. 
This gives a simple pictorial way to represent the flavor content 
of the ensemble by plotting the tip of the projection of
$\Pvec(\omega,r)$ on the ${\bf e}_{3}$--${\bf e}_{8}$ 
plane. \footnote{Note that probability conservation in this 
representation corresponds to the theorem that the sum of the 
lengths of perpendiculars dropped from any point inside an 
equilateral triangle to the three sides is a constant.}

For the mass term in the Hamiltonian, we have
\barr
H_{vac}(\omega,h)&=&h\omega\left(\frac{1}{3}\;
{\rm B}_0\;I +\frac{1}{2}\;\Bvec \cdot {\bm \Lambda}\right)~,
\label{hobvec}
\earr
where
\barr
h~\Bvec&=&\epsilon c_{13}\sin2\theta_{12}\;\hat{\bf e}_{1}+
\bigg(s_{13}^2 - \epsilon(c_{12}^2-c_{13}^2s_{12}^2) \bigg)\;
\hat{\bf e}_{3} + 
(1-\epsilon s_{12}^2)\sin2\theta_{13}\;\hat{\bf e}_{4}\nonumber\\
&&-\epsilon s_{13}\sin2\theta_{12}\;\hat{\bf e}_{6}+ 
\bigg((-2+\epsilon)(3c_{13}^2-1)+3\epsilon s_{13}^2(2c_{12}^2-1)
\bigg)/(2\sqrt{3})\;\hat{\bf e}_{8}~.
\label{Bvecdefn}
\earr
Note that $\omega$ for neutrinos is always positive in this 
convention, and the negative sign of $\dmsq_{13}$ for 
inverted hierarchy is absorbed into $\Bvec$. 
The terms involving $\epsilon=\dmsq_{12}/\dmsq_{13}$ 
arise from the mixing of the third flavor, 
and the three-flavor effects enter through them. 
The sign of $\epsilon$ is positive if the mass hierarchy is normal 
($\dmsq_{13}>0$) and negative otherwise. 
This, along with the overall sign due to $h$, guarantees that the 
contributions from $\dmsq_{12}$ always have the same sign. 
Note that ${\rm B}_{2},\;{\rm B}_{5},\;{\rm B}_{7}$ 
vanish in the absence of $CP$-violation. 

The MSW potential defined in Eq. (\ref{msw-potential}) may be represented as
\barr
V(r)&=&\lambda(r)\left(\frac{1}{3}\;{\rm L}_0\;I +\frac{1}{2}\;\Lvec
\cdot {\bm \Lambda}\right)~,
\label{lambdadefn}
\earr
where $\lambda(r)=\sqrt{2}G_{F}\;n_{e^-}(r)$. 
The vector $\Lvec$ parameterizes the effect of background electrons, 
and is given by
\beq
\Lvec=\hat{\bf e}_{3}+ \hat{\bf e}_{8} / \sqrt{3}~.
\label{Lvecdefn}
\eeq
The $H_{\nu\nu}(r)$ term defined in Eq. (\ref{hnunu-r}) can also be simply written as
\barr
H_{\nu\nu}(r)&=&\mu(r)\left(\frac{1}{3}\;{\rm D}_0\;I +
\frac{1}{2}\;\Dvec(r) \cdot {\bm \Lambda}\right)~,
\earr
where $\Dvec(r)$ is defined as
\beq
\Dvec(r)=\int d\omega\;f(\omega)\;\Pvec(\omega,r)\;sgn(\omega)~.
\label{Dvecdefn}
\eeq
In the next section, we shall express the evolution equation, i.e. 
Eq.~(\ref{sageom})
in terms of the Bloch vectors 
$\Pvec(\omega,r), \Bvec(\omega,h), \Lvec$ and $\Dvec(r)$.

\subsection{Generalized Gyroscope Equations}
\label{gyroscope}

We have expressed our problem in terms of the eight-dimensional 
Bloch vectors, and now we shall see that the equations of motion 
formally resemble the equations of a gyroscope. 
To make this apparent, we define $\times$ as a generalized 
``cross product'' \cite{kim-kim-sze}
with $f_{abc}$ as the structure constants, 
instead of the usual $\epsilon_{abc}$ that appears in the 
two-flavor approximation, e.g.
\beq
{\Bvec}\times {\Pvec} \equiv 
\sum_{a,b=1}^{8}f_{abc}{\rm B}_{a}{\rm P}_{b}{\;\hat{\bf e}_{c}}~.
\label{cross-product}
\eeq
This makes it possible to write the equations of motion, i.e. Eq.~(\ref{sageom}), compactly as
\beq
\dot{\Pvec}(\omega,r)=\bigg(\omega \Bvec+ \lambda(r)\Lvec + 
\mu(r)\Dvec(r)\bigg)\times \Pvec(\omega,r)~ \equiv
{\rm \bf H}(\omega, r) \times \Pvec(\omega,r)~,
\label{gyroeom}
\eeq
where $\Pvec(\omega,r),\;\Bvec,\;\Lvec,\;\Dvec(r)$ 
are defined in Eqs. (\ref{polvec}), (\ref{Bvecdefn}), (\ref{Lvecdefn}) and (\ref{Dvecdefn}) respectively. 
The couplings $\omega$, $\mu(r)$ and $\lambda(r)$ are defined in  
Eqs. (\ref{omegadefn}), (\ref{dilfac}) and (\ref{lambdadefn}) respectively. 
Equation (\ref{gyroeom}) resembles the equation of a spin in 
an external magnetic field, or equivalently, that of a gyroscope. 
We must remember that this similarity is purely formal, 
because unlike in the two-flavor case, we cannot write an 
arbitrary Bloch vector as a linear combination of two 
Bloch vectors and their cross product.  
We shall show in Sec.~\ref{factorization} that under 
certain approximations these generalized gyrosope equations 
can be given a geometrical interpretation.

The effects of the matter term $\lambda(r) \Lvec$ in 
Eq. (\ref{gyroeom}) can be rotated away by going to the
interaction frame as described in Eq. (\ref{simtransf}),
where a matrix $A$ becomes $A^{int} = O A O^{-1}$.
In order to determine the Bloch vector corresponding to
$A^{int}$, we equate
\beq
\frac{{\rm A}_0}{3} I + \sum_{a=1}^{8} \frac{{\rm A}_a \Lambda_a}{2} 
= O A O^{-1} \; .
\eeq
Multiplying both sides by $\Lambda_a$ and taking trace,
we get
\beq
{\rm A}_a^{int} = {\rm Tr}(\Lambda_a O A O^{-1}) \; ,
\label{int-component}
\eeq
where we have used ${\rm Tr}(\Lambda_a \Lambda_b)= 2\delta_{ab}$.
In particular, the Bloch vector $\Bvec^{int}$ may be written
using Eqs. (\ref{O-def}) and (\ref{int-component}) as
\barr
\Bvec^{int}(r)&=&
{\rm B}_1 \cos \zeta(r) ~\hat{\bf e}_{1} + 
{\rm B}_1 \sin \zeta(r)~ \hat{\bf e}_{2} + 
{\rm B}_3 ~\hat{\bf e}_{3} \nonumber\\
&&+ {\rm B}_4 \cos \zeta(r)~ \hat{\bf e}_{4} + 
{\rm B}_4 \sin \zeta(r)~\hat{\bf e}_{5} 
+ {\rm B}_6\;\hat{\bf e}_{6}+ {\rm B}_7\;\hat{\bf e}_{7}+ 
{\rm B}_8\;\hat{\bf e}_{8}~,
\label{bint}
\earr
where $\zeta(r) = \int_0^r V(r') dr'$.
In dense matter, ${\rm B}^{int}_a(r)$ oscillates rapidly with 
the frequency $\sim V(r)$, mimicking a suppression in the 
relevant mixing angles as in the two-flavor case 
\cite{hannestad-raffelt-sigl-wong-0608095}.

We also define the ``signed'' and ``unsigned''
$n^{th}$ moments (with $n\geq 0$) of $\Pvec(\omega,r)$ as
\barr
\Dvec^{(n)}(r)&=&\int d\omega \;\omega^{n}\;f(\omega)\;
\Pvec(\omega,r)\;sgn(\omega)~, \\
\Svec^{(n)}(r)&=&\int d\omega\; \omega^{n}\;f(\omega)\;
\Pvec(\omega,r)~.
\label{moments-def}
\earr
Note that $\Dvec^{(0)}$ is same as $\Dvec$, and we will therefore refer to $\Svec^{(0)}$ as $\Svec$. The evolution of these 
moments are governed by
\barr
\dot\Dvec^{(n)}(r)&=&\Bvec\times\Dvec^{(n+1)}(r)+ 
\bigg(\lambda(r)\Lvec + \mu(r)\Dvec(r) \bigg)
\times\Dvec^{(n)}(r)~, ~\label{dndot}\\
\dot\Svec^{(n)}(r)&=&\Bvec\times\Svec^{(n+1)}(r)+ 
\bigg(\lambda(r)\Lvec + \mu(r)\Dvec(r) \bigg)
\times\Svec^{(n)}(r)~\label{sndot}.
\earr
We see that the higher moments turn up in equations of motion 
the lower moments. 
If we take the dot product of Eq. (\ref{dndot}) 
with $\Dvec^{(n)}(r)$, 
and of Eq. (\ref{sndot}) with $\Svec^{(n)}(r)$, we get
\barr
\partial_{r}{|\Dvec^{(n)}(r)|^2}&=&\Dvec^{(n)}(r) \cdot 
\Bvec\times\Dvec^{(n+1)}(r)~,\nonumber\\
\partial_{r}{|\Svec^{(n)}(r)|^2}&=&\Svec^{(n)}(r) \cdot 
\Bvec\times\Svec^{(n+1)}(r)~.
\earr
The above dependence of the moments on $r$ implies that there is likely to be 
a redistribution of flavor as a function of $\omega$. 
It will be interesting to investigate if these moment equations 
can be used to predict the nature of the redistribution of 
flavor spectra.

\subsection{Heavy-Light factorization of dynamics}
\label{factorization}

The three-flavor dynamics in the traditional matter-driven scenario 
can be factorized into the so-called ``heavy'' and  ``light'' 
MSW resonances that occur at densities corresponding to 
$\msqa$ and $\msqs$ respectively. 
Appropriate combination of the effective two-flavor dynamics
in these two sectors
approximates the three-flavor result reasonably well. 
We now proceed to illustrate a similar simplification for 
collective effects as well
Let us first introduce the notion of ``heavy'' and ``light'' 
subspaces of the Bloch-sphere.
In the ${\mathbbm K}$-${\mathbbm Q}$ decomposition shown in 
Eq. (\ref{Key}), the vectors contained in ${\mathbbm K}^{ey}$ are
termed ``heavy'' ($H$) whereas those contained in ${\mathbbm Q}^{ey}$ are
termed ``light'' ($L$).
A general vector $\Xvec$ may be decomposed as
\beq
\Xvec=\Xvec^{H} + \Xvec^{L}~.
\eeq
In particular, $\Bvec$ in Eq. (\ref{Bvecdefn})
may be expressed as $\Bvec=\Bvec^{H}+ \Bvec^{L}$, with
\barr
h \Bvec^{H}&=&
\bigg(s_{13}^2 - \epsilon(c_{12}^2-c_{13}^2s_{12}^2) \bigg)\;
\hat{\bf e}_{3} + 
(1-\epsilon s_{12}^2)\sin2\theta_{13}\;\hat{\bf e}_{4} \nonumber \\
& & +
\bigg((-2+\epsilon)(3c_{13}^2-1)+3\epsilon s_{13}^2(2c_{12}^2-1)
\bigg)/(2\sqrt{3})\;\hat{\bf e}_{8} \; , \nonumber \\
h \Bvec^{L}&=& \epsilon c_{13}\sin2\theta_{12}\;\hat{\bf e}_{1}
-\epsilon s_{13}\sin2\theta_{12}\;\hat{\bf e}_{6} \; . 
\earr
The component $\Bvec^{H}$ appears primarily due to 
$\dmsq_{13}$, and the other component $\Bvec^{L}$ vanishes 
if $\epsilon=0$. 
Note that for two-flavors, or equivalently in the 
$\epsilon=0$ limit, $\Bvec$ is completely contained in 
${\mathbbm K}^{ey}$. 
Now, note the following structure in the equations of motion 
of a polarization vector:
\barr
\dot{\Pvec}^{H}(\omega, r)&=& {\rm \bf H}^{H}(\omega,r) 
\times\Pvec^{H} (\omega,r)
+ {\rm \bf H}^L (\omega,r)\times\Pvec^{L}(\omega,r)~,\label {pdotH}\\
\dot{\Pvec}^{L}(\omega, r)&=& {\rm \bf H}^{L}(\omega,r) 
\times\Pvec^{H} (\omega,r)
+ {\rm \bf H}^H (\omega,r)\times\Pvec^{L}(\omega,r)~.
\label {pdotL}
\earr
It is clear from the above set of equations that if $\epsilon=0$ 
and one begins with $\Pvec$ contained in ${\mathbbm K}^{ey}$, 
then $\Pvec$ always remains in ${\mathbbm K}^{ey}$, i.e. 
$\Pvec^{L}(\omega,r)=0$ identically. 
To investigate this case more closely, we write Eq. (\ref{pdotH}) 
for each component of $\Pvec^{H}$ 
as {\footnote{In the following sections, the dependence of the
Bloch vectors and the parameters on $\omega$ and $r$ is implicit.}
\barr
\dot{\rm P}_{3}&=& {\rm H}_4 {\rm P}_5 - {\rm H}_5 {\rm P}_4 \; ,\\
\dot{\rm P}_{4}&=& {\rm H}_5 {\rm P}_3 - {\rm H}_3 {\rm P}_5 + 
\sqrt{3} ({\rm H}_5 {\rm P}_8 - {\rm H}_8 {\rm P}_5) \; ,\\
\dot{\rm P}_{5}&=& {\rm H}_3 {\rm P}_4 - {\rm H}_4 {\rm P}_3 + 
\sqrt{3} ({\rm H}_8 {\rm P}_4 - {\rm H}_4 {\rm P}_8) \; ,\\
\dot{\rm P}_{8}&=& \sqrt{3} ({\rm H}_4 {\rm P}_5 - 
{\rm H}_5 {\rm P}_4) \; .
\earr
Note that $\dot{\rm P}_{8}=\sqrt{3}\;\dot{\rm P}_{3}$. 
This suggests that we could rotate our coordinates in the 
${\bf e}_{3}$--${\bf e}_{8}$ plane by $-2 \pi/3$, 
so that $\widetilde{\rm P}_8$ in the rotated frame becomes 
a constant of motion. 
While going to the rotated frame, the components
${\rm X}_3$ and ${\rm X}_8$ of any Bloch vector 
$\Xvec$ transform as
\beq
\left( \begin{array}{c}
\widetilde{\rm X}_3 \\ \widetilde{\rm X}_8 
\end{array} \right) =
\left(\begin{array}{cc}
-1/2&-\sqrt{3}/2\\
\sqrt{3}/2&-1/2
\end{array}\right)
\left( \begin{array}{c}
{\rm X}_3 \\ {\rm X}_8 
\end{array} \right) \; .
\label{def-R}
\eeq
The other components remain unchanged.

This leads to the following simplified equations of motion 
for the two-flavor case:
\barr
\dot{\widetilde{\rm P}}_{3}&=& -2({\rm H}_4 {\rm P}_5 - {\rm H}_5 
{\rm P}_4) \; , \label{tilde-eqns1}\\
\dot{\widetilde{\rm P}}_{4}&=& -2 ({\rm H}_5 \widetilde{\rm P}_3 - 
\widetilde{\rm H}_3 {\rm P}_5) \; , \label{tilde-eqns2}\\
\dot{\widetilde{\rm P}}_{5}&=& -2(\widetilde{\rm H}_3 {\rm P}_4 - 
{\rm H}_4 \widetilde{\rm P}_3) \; ,\label{tilde-eqns3}\\
\dot{\widetilde{\rm P}}_{8}&=& 0 \; .\label{tilde-eqns4}
\earr
This is the two-flavor limit, where the state $\nu_x$ does not 
participate in the evolution. 
This is a consequence of all the polarization vectors initially
being contained in ${\mathbbm K}^{ey}$. 
The rotated ``tilde'' frame can therefore be 
called as the ``$e-y$'' frame.

The Eqs. (\ref{tilde-eqns1}), (\ref{tilde-eqns2}) and (\ref{tilde-eqns3}) can be simply written as
\beq
\dot{\rm \bf P}^{ey} = {\rm \bf H}^{ey} \times \Pvec^{ey} \; ,
\eeq
where the ``$\times$'' can now be taken to be the usual cross product 
in a three-dimensional space spanned by 
$\{ {\bf e}_3^{ey},\;{\bf e}_{4},\;{\bf e}_{5} \}$. 
This clearly exhibits the ``gyration'' of $\Pvec$ 
about ${\rm \bf H}$, while the component of $\Pvec$ 
along ${\bf e}_{8}^{ey}$ remains constant. 
The projection of $\Pvec$ changes only along ${\bf e}_3^{ey}$, 
which corresponds to $\nu_e \leftrightarrow \nu_y$ flavor
conversions.
The problem is thus reduced to the two-flavor limit, 
for which analytical solutions have been discussed in literature
\cite{pastor-raffelt-semikoz-0109035,hannestad-raffelt-sigl-wong-0608095,raffelt-smirnov-0705.1830}.

In the two-flavor limit, it is observed that there are three 
qualitatively different kinds of motion of the polarization vector 
in the flavor space. 
The most familiar case is oscillations in vacuum/matter, where the neutrino-antineutrino density 
is small ($\mu\ll\omega$) and each $\Pvec(\omega)$ precesses 
about $\Bvec$ with frequency $\omega$. 
The other extreme is when the neutrino-antineutrino density is 
very large ($\mu\gg\omega$). 
In such a situation, all $\Pvec(\omega)$ remain tightly bound 
together and precess with the average $\omega$ of the ensemble, 
giving rise to synchronized oscillations. 
The intermediate regime ($\mu \gsim \omega$) is when the 
$\Pvec(\omega)$ remain bound together to a large extent, 
but have a tendency to relax to the state that has the lowest 
energy. 
The system is analogous to a pendulum/gyroscope that tries to 
relax to its vertically downward state, 
whatever state one might start in. 
This motion is called bipolar oscillation.

The motion changes qualitatively and quantitatively with the 
inclusion of the third flavor. There are two kinds of 
contribution due to the inclusion of the third flavor. 
First, we have some extra contributions to $\Bvec^{H}$ 
that depend on $\epsilon$, which changes the effective 
values of $\omega$ and $\theta_{13}$. 
These do not change the motion qualitatively. 
The second type of contribution is more interesting. 
It is due to the excursions of the polarization vectors 
into the ${\mathbbm Q}$ subspace under the influence of 
$\Bvec^{L}$. 
In particular, the length of $\Pvec^{H}$ is not preserved anymore. 
To see this clearly, we take the dot product of $\Pvec^{H}$ 
with Eq. (\ref{pdotH}) and that of $\Pvec^{L}$ 
with Eq. (\ref{pdotL}) to get
\beq
{|\dot{\Pvec}^{H}|^2}/2=-{|\dot{\Pvec}^{L}|^2}/2=
\Pvec^{H} \cdot {\rm \bf H}^{L}  \times\Pvec^{L}
~.
\label{modpvariation}
\eeq
We can clearly see that $|\Pvec^{H}|$, which was a conserved 
quantity in the two-flavor case, no longer remains so. 
The non-conservation is proportional to $|{\rm \bf H}^L|$ and 
$|\Pvec^{L}|$, both of which go to zero in the two-flavor limit. 
The addition of the third flavor makes the 
motion of the projection of $\Pvec$ in the ${\bf e}_3$--${\bf e}_8$ 
plane fairly complicated in general, 
and we shall study it in some interesting regimes in
Sec.~\ref{toy-cases}.

\subsection{The three-flavor solution}
\label{three-flavor}

In this section we extend the method presented in the last section 
to include the leading corrections due to the mixing of the 
third flavor. 
Let us illustrate our prescription in the vacuum limit, where
the matter effects as well as the collective effects are
neglected. 
The prescription will later be easily generalized to finite
matter densities and significant neutrino-neutrino
interactions.

From Eq. (\ref{Bvecdefn}), the Bloch vector $\Bvec$ may be
decomposed as
\beq
\omega \Bvec = h \omega \Bvec^{(1)} + h \epsilon \omega \Bvec^{(2)}
+ h \epsilon \omega s_{13} \Bvec^{(3)}
\label{b1b2b3}
\eeq
with
\barr
\Bvec^{(1)} & = & s_{13}^2 ~\hat{\bf e}_3 
 -2(3c_{13}^2-1)/(2\sqrt{3}) \;\hat{\bf e}_{8} 
+ (1-\epsilon s_{12}^2)\sin2\theta_{13}\;\hat{\bf e}_{4} \; , \\
\Bvec^{(2)} & = &
-(c_{12}^2-c_{13}^2s_{12}^2)\; \hat{\bf e}_{3} + 
(3c_{13}^2-1)/(2\sqrt{3})\;\hat{\bf e}_{8} +
c_{13}\sin2\theta_{12}\;\hat{\bf e}_{1} \; , \\
\Bvec^{(3)} & = &
3 s_{13}(2c_{12}^2-1)/(2\sqrt{3})\;\hat{\bf e}_{8}
- \sin2\theta_{12}\;\hat{\bf e}_{6} \; .
\label{b1b2b3-def}
\earr
Note that $\Bvec^{(1)}$ lies completely in ${\mathbbm K}^{ey}$,
$\Bvec^{(2)}$ in ${\mathbbm K}^{ex}$, and
$\Bvec^{(3)}$ in ${\mathbbm K}^{xy}$.

In Fig.~\ref{3-coordinates}, we show three coordinate frames
$e-x, e-y$ and $x-y$ in the ${\bf e}_3$--${\bf e}_8$ plane.
These frames are defined such that, if ${\textsf P}$ is the
projection of $\Pvec$ in the ${\bf e}_3$--${\bf e}_8$ plane,
the components $\Bvec^{(1)}, \Bvec^{(2)}, \Bvec^{(3)}$ 
in Eq. (\ref{b1b2b3-def})
separately cause ${\textsf P}$ to move along 
${\bf e}_3^{ey}, {\bf e}_3^{ex}, {\bf e}_3^{xy}$ respectively.
In order to reduce the motions due to $\Bvec^{(1)},\Bvec^{(2)},
\Bvec^{(3)}$
separately to two flavor problems as in Sec.~\ref{factorization},
we write 
\beq
\Bvec^{ey} = {\textsf R} \Bvec^{(1)} \; , \qquad
\Bvec^{ex} =  \Bvec^{(2)} \; , \qquad
\Bvec^{xy} = {\textsf R}^2 \Bvec^{(3)} \; , 
\eeq
where ${\textsf R}$ is the rotation matrix in Eq. (\ref{def-R})
that rotates the $\Xvec_3$ and $\Xvec_8$ components of a 
Bloch vector in the ${\bf e}_3$--${\bf e}_8$
plane by $-2 \pi/3$. 
The vectors $\Bvec^{ey},\Bvec^{ex},\Bvec^{xy}$ are then simply
$\Bvec^{(1)},\Bvec^{(2)},\Bvec^{(3)}$ 
in the frames $e-y,e-x,x-y$ respectively.
\begin{figure}
\epsfig{file=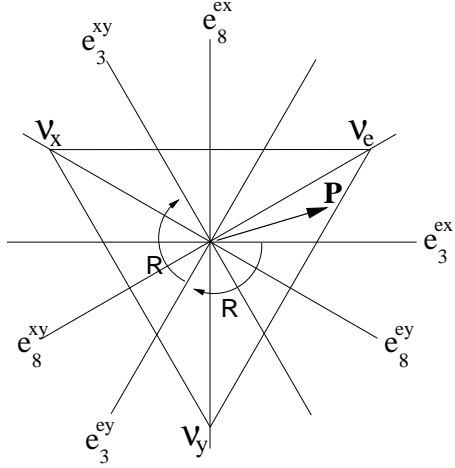,width=6.0cm}
\caption{Useful coordinate choices on ${\bf e}_3$--${\bf e}_8$ plane}
\label{3-coordinates}
\end{figure}
We can then write Eq. (\ref{b1b2b3}) as
\beq
\omega \Bvec = 
\omega^{ey} {\textsf R}^{-1}~ \Bvec^{ey} +
\omega^{ex} \Bvec^{ex} +
\omega^{xy} {\textsf R}^{-2}~ \Bvec^{xy} \; ,
\eeq
with the ``frequencies'' defined as
\beq
\omega^{ey} = h \omega  \; \qquad
\omega^{ex} = h  \epsilon \omega \; \qquad
\omega^{xy} =  h \epsilon \omega s_{13} \sin 2\theta_{12}\;,
\label{omegas}
\eeq
and the ``magnetic fields'' as
\barr
\Bvec^{ey} & = &  \cos 2 \theta_{13} ~\hat{\bf e}_3 +
 (1-\epsilon s_{12}^2)\sin2\theta_{13}\;\hat{\bf e}_{\perp}^{ey} - (1-3 s^2_{13})/(\sqrt{3}) ~\hat{\bf e}_{8} \; \\
\Bvec^{ex} & = & -(c_{12}^2-c_{13}^2s_{12}^2)\; \hat{\bf e}_{3} + 
c_{13}\sin2\theta_{12}\;\hat{\bf e}_{\perp}^{ex}  +
(3c_{13}^2-1)/(2\sqrt{3})\;\hat{\bf e}_{8}  \; , \\
\Bvec^{xy} & = & - \hat{\bf e}_{\perp}^{xy} 
- \sqrt{3} s_{13} \cos 2 \theta_{12}/(4 \sin 2 \theta_{12}) 
\;\hat{\bf e}_{8} \; .
\label{balphabeta-def}
\earr
The vectors ${\bf e}_4, {\bf e}_1, {\bf e}_6$ are the 
directions transverse to the ${\bf e}_3$--${\bf e}_8$ plane
that are relevant in the three frames, and can be written as 
${\bf e}_{\perp}^{ey}$, ${\bf e}_{\perp}^{ex}$, ${\bf e}_{\perp}^{xy}$ respectively.
The $B^{\alpha\beta}$ are normalized such that 
$|B^{\alpha\beta}_3|^2 + |B^{\alpha\beta}_\perp|^2 
= 1 + {\cal O}(\epsilon, s_{13}^2)$.
The separate motion due to each $B^{\alpha\beta}$ is then 
a precession about 
$B^{\alpha\beta}_3 \hat{\bf e}_3 + B^{\alpha\beta}_\perp
\hat{\bf e}_\perp$ with a frequency $\omega^{\alpha\beta}$, 
where the half-angle of the cone is given by
$\tan\theta^{\alpha\beta} =  |B^{\alpha\beta}_\perp/
 B^{\alpha\beta}_3|$.

The net motion of the polarization vector may  be
interpreted as the combination of two-flavor precessions 
about ${\bf e}_\perp^{ey}, {\bf e}_\perp^{ex}$ 
and ${\bf e}_\perp^{xy}$ respectively.
It can immediately be seen from Eq. (\ref{omegas}) that
\beq
|\omega^{ey}| \gg |\omega^{ex}| \gg |\omega^{xy}| \; ,
\label{w-hierarchy}
\eeq
i.e. the precession frequencies are hierarchical.
Therefore, the motion due to slower frequencies may be neglected 
over short time scales.
More precisely, if we coarse-grain the equation of 
motion Eq. (\ref{gyroeom}) in $r$ over scales corresponding to 
$\omega^{ey}$, the effects of $\omega^{ex}$ and $\omega^{xy}$
are negligible. 
The slowest variation in the solution is due to $\omega^{xy}$,
which modulates the faster motion due to $\omega^{ex}$,
which in turn modulates the motion at still shorter scales
due to $\omega^{ey}$.

Let us denote the evolution of ${\textsf P}(r)$ under the action 
of $\Bvec^{ey},\;\Bvec^{ex},\;\Bvec^{xy}$ by the operators 
${\textsf S}^{ey}(r),\; {\textsf S}^{ex}(r),\;{\textsf S}^{xy}(r)$ 
respectively. 
As long as the condition in Eq. (\ref{w-hierarchy}) is valid, we can write
\beq
{\textsf P}(r)= {\textsf S}^{ey}(r) \; {\textsf S}^{ex}(r)
\;{\textsf S}^{xy}(r) \;{\textsf P}(0)~,
\label{fullsoln}
\eeq
where the evolution operators are of the form
\barr
{\textsf S}^{ey}(r)&=& {\textsf R}^{-1} \left(\begin{array}{cc}
{\eta}(\omega^{ey}, \theta^{ey},\mu,r)&0\\0&1\end{array}\right)
{\textsf R}~,\\
{\textsf S}^{ex}(r)&=&\left(\begin{array}{cc}
{\eta}(\omega^{ex}, \theta^{ex}, \mu,r)&0\\0&1\end{array}\right)~,\\
{\textsf S}^{xy}(r)&=& {\textsf R}^{-2} \left(\begin{array}{cc}
{\eta}(\omega^{xy}, \theta^{xy}, \mu,r)&0\\0&1\end{array}\right)
{\textsf R}^{2} ~.
\label{s-matrices}
\earr
Here $\eta(\omega^{\alpha\beta},\theta^{\alpha\beta},\mu,r)$ 
are the 
evolution functions that can be calculated in a two-flavor 
approximation using the results in previous literature. 
In general, the frequencies of these evolution functions
are determined by $\omega^{\alpha\beta}$s and the amplitudes are
determined by the effective mixing angle $\theta^{\alpha\beta}$s. 
Each evolution operator ${\textsf S}^{\alpha\beta}$ takes the 
state ${\textsf P}$ to the respective $\alpha-\beta$ frame
in which ${\rm P}_8^{\alpha\beta}$ stays constant and 
${\rm P}_3^{\alpha\beta}$ undergoes precession, and brings
${\textsf P}$ back to the ${\bf e}_3$--${\bf e}_8$
frame after precession.
Note the matrices ${\textsf S}^{\alpha\beta}$ are not unitary.
The order in which they are operated should be such that 
the slower oscillations effectively act like an amplitude 
modulation for the faster oscillations. 

It is easy to calculate $\rho_{\nue\nue}$ using Eq. (\ref{rhoeefinal}) as
\beq
\rho_{\nue\nue}(r)=\frac{1}{3}+
\frac{{\textsf P}(r) \cdot {\bf e}_{e}}{2}
= \frac{1}{3}+\frac{1}{\sqrt{3}}
\left(-\frac{\sqrt{3}}{2} {\rm P}_{3}^{ey}(r) + 
\frac{1}{2} {\rm P}_{8}^{ey} \;(r)\right)~,
\label{rhoeeP}
\eeq
where ${\textsf P}(r)$ is given by Eq. (\ref{fullsoln}),
and ${\rm P}_3^{ey}, {\rm P}_8^{ey}$ are components
along ${\bf e}_{3}^{ey}$ and ${\bf e}_{8}^{ey}$ respectively.
If neglect effects of the slowest frequency $\omega^{xy}$,
the expressions for ${\rm P}_{3}^{ey}(r)$ and ${\rm P}_{8}^{ey}(r)$ 
may be written as
\barr
{\rm P}_{3}^{ey} \;(r)&=& \eta(\omega^{ey}, \theta^{ey}, \mu, r) 
\left(-\frac{1}{2} \eta(\omega^{ex}, \theta^{ex}, \mu, r) {\rm P}_3(0)
- \frac{\sqrt{3}}{2} {\rm P}_8(0) \right)~,
\label{3flavp3}\\
{\rm P}_{8}^{ey} \;(r)&=&
\left(+\frac{\sqrt{3}}{2}\eta(\omega^{ex}, \theta^{ex}, \mu, r) 
{\rm P}_3(0)
- \frac{1}{2} {\rm P}_8(0) \right)~.
\label{3flavp8}
\earr

In the presence of ordinary matter and when the collective effects
may be neglected, the same prescription stays valid, 
simply with the replacements
\beq
\hat{\bf e}_1 \to \cos \zeta(r) \; \hat{\bf e}_1 +
\sin \zeta(r) \;  \hat{\bf e}_2  \; , \quad
\hat{\bf e}_4 \to  \cos \zeta(r) \; \hat{\bf e}_4 +
\sin \zeta(r) \; \hat{\bf e}_5  \; 
\label{replacements}
\eeq
with $\zeta(r) = \int_0^r V(r') dr'$.
It may be seen from Eq. (\ref{bint}) that these replacements 
take $\Bvec$ to $\Bvec^{int}$, so that the effect of MSW 
is taken into account by going to the interaction frame.
As observed in Sec.~\ref{gyroscope}, fast oscillations
with a frequency $\sim V(r)$ will average out the sinusoidal
terms, thus decreasing the contribution from the
transverse components of $\Bvec^{(int)\alpha\beta}$.

When the collective effects dominate, since the collective
potential $H_{\nu\nu}(r)$ in Eq. (\ref{sageom}) is independent
of energy, neutrinos of all energies precess with a 
common frequency in all the two-flavor subspaces.
The motion is therefore similar to the vacuum case 
discussed above, with the replacement
$\omega \to \langle \omega \rangle$ as given in
Sec.~\ref{synchronized}.

We have thus completed our program of expressing three-flavor 
effects purely in terms of two-flavor effects. 
The $r$-dependent functions $\eta(r)$ are known analytically for 
oscillations in vacuum and for synchronized oscillations, 
where we can explicitly check our ansatz. 

In the case of bipolar oscillations, the situation is more 
complicated since these are not sinusoidal oscillations,
rather $\Pvec$ remains almost static for a period of time 
and swings through the lowest energy state in a rapid burst. 
As a result, the fast- or slow-ness of bipolar oscillations
as compared to the other precessions is time dependent.
We therefore can obtain a qualitative understanding of 
bipolar oscillations in the three neutrino framework, but
only a heuristic form of the analytic solution.

\section{Flavor Conversion Mechanisms in three-flavor formalism}
\label{toy-cases}

In this section we illustrate the three-flavor effects
in some simple examples, where we take constant matter
density and box-spectra for neutrinos and antineutrinos.
We explain the three-flavor features therein analytically 
using the ``${\bf e}_3$--${\bf e}_8$'' 
triangle diagrams.
The insights gained thereby will allow us to understand
the more complicated flavor conversions in realistic
supernova simulations in Sec.~\ref{full-sn}.
For our numerical evaluations in this section, 
we fix $|\msqa|=2.5\times10^{-3}~{\rm eV}^2$ and
$\theta_{12}=0.6$. We also choose a box-spectrum for the the neutrino flux
i.e. constant over the energy range $E =(1$--$51)$ MeV, and zero elsewhere.

\subsection{Vacuum and MSW oscillations}
\label{vac-msw}
We start with looking at neutrino oscillations in vacuum/matter, 
with no collective effects. Although this situation has been 
analyzed in literature in great detail, 
we illustrate it here in order to 
familiarize the reader with the analysis in terms of 
${\rm P}_3^{ey}, {\rm P}_8^{ey}$ and the 
``${\bf e}_3$--${\bf e}_8$'' triangle.
This triangle, shown in
Fig.~\ref{fig:vac-msw}, helps in understanding the
three-neutrino features of flavor conversions.
The projection of $\Pvec$ on the ${\bf e}_3$--${\bf e}_8$ plane
represents the flavor content, the allowed region being an
equilateral triangle.
The three vertices of the triangle represent the three states 
$\nue$, $\nux$ and $\nu_y$ (anticlockwise, from top right).
States that lie on the edges connecting them are admixtures of 
only those two flavors.
The interior of the triangle represents states that are 
admixtures of all three flavors. 
Quantitatively, for any point on the triangle, 
the fraction of the neutrinos 
in flavor $\alpha$ is proportional to its distance from the 
edge opposite to the $\nu_\alpha$ vertex,
as shown in Eq.~(\ref{rhoeefinal}).

In Fig.~\ref{fig:vac-msw}, we show the quantities 
${\rm P}_3^{ey}, {\rm P}_8^{ey}$ and $\rho_{\nu_e \nu_e}$
as functions of the radial coordinate $r$. For illustration, we start with a pure $\nu_e$ flavor, 
which corresponds to $({\rm P}_3^{ey}, {\rm P}_8^{ey}) = (-1, 1/\sqrt{3})$.
The following observations may be made from the figure.

\begin{figure}
\parbox{9cm}{
\epsfig{file=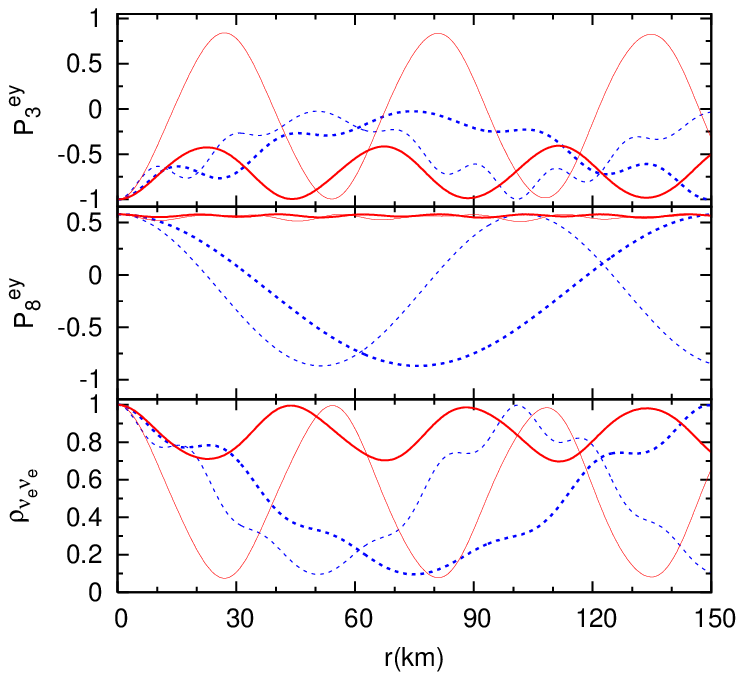,width=8.5cm}
}
\parbox{7cm}{
\epsfig{file=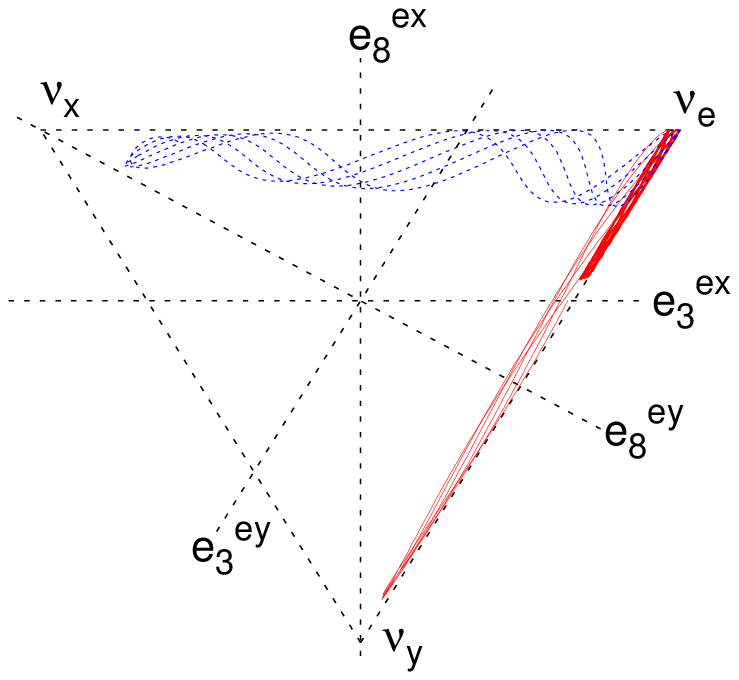,width=7cm}
}
\caption{Neutrino oscillations for $E = 20.0$ MeV and $29.6$ MeV 
(thin and thick lines respectively). 
To emphasize the nature the oscillations, 
we choose $\epsilon=1/5.1$ and  $\theta_{13}=0.2$.
Oscillations in vacuum and matter are shown by
dotted (blue) and undotted (red) lines respectively. 
For matter, we choose normal hierarchy and  $\lambda= 0.3~{\rm km}^{-1}$.
}
\label{fig:vac-msw}
\end{figure}

\begin{itemize}

\item The oscillation frequencies depend on the neutrino energy.
However in the triangle diagram, the locus of ${\textsf P}$
for all energies is identical for oscillations in vacuum 
(therefore, the thin and thick lines overlap). 
Different energies 
travel along this orbit at different, but constant speeds proportional to $1/E$. 
In matter, the mixing angle begins to depend on the 
energy and thus the orbits are different for different energies.

\item The flavour evolution has two main frequency components, 
The fast oscillations with frequency 
$\omega=\msqa/(2p)$ and the slower
ones with frequency $\epsilon \omega = \msqs/(2p)$.

\item If $\omega$ and $\epsilon \omega$ were commensurate, 
the orbits in the triangle would be closed curves. 
However, that is a fine-tuned situation.
In general, if $\epsilon$ is not rational,
the orbits do not close, but drift parallel to 
themselves periodically. Indeed, the orbits are analogous to 
the well-known Lissajous figures.

\item ${\rm P}_8^{ey}$ only has slow modes corresponding to 
the frequency $\epsilon \omega$. 
These slow oscillations modulate the amplitude
of the upper envelope of $|{\rm P}_3^{ey}|$ because the
maximum value that $|{\rm P}_3^{ey}|$ can take is reduced 
when ${\rm P}_8^{ey}$ deviates from its maximum value of
$1/\sqrt{3}$. The above can be clearly seen from the
triangle diagram.

\item ${\rm P}_3^{ey}$ oscillations involve both frequencies, 
$\omega$ and $\epsilon \omega$. 
The maximum deviation of ${\rm P}_3^{ey}$
from unity is governed by the amplitude of modulation of its 
upper envelop (which depends on $\sin^2 2 \theta_{13}$) 
and the amplitude of faster oscillations superimposed on it 
(which depends on $\sin^2 2 \theta_{12}$).

\item In the two-flavor limit we ignore the mixing 
with $\nu_x$, and as a result ${\rm P}_8^{ey}$ remains constant. 
In the triangle, this corresponds to the motion being
confined to a line parallel to the ${\bf e}_3^{ey}$ axis.
Indeed, the effect of the third flavor is to extend the
motion of ${\textsf P}$ to the entire triangle, as opposed
to only along a line. 
The deviation of ${\textsf P}$ from this line quantifies 
the extent of three-flavor effects.

\item
The amplitude of oscillations can be read off from the triangle as the 
extent of the orbit along the $\nu_e$--$\nu_y$ edge 
($2 \sin^2 2\theta_{13}$) 
and along the $\nu_e$--$\nu_x$ edge 
($2\sin^2 2\theta_{12}$). 

\item In the presence of matter, mixing angles are suppressed
or enhanced depending on the energy and matter density.
For $\lambda \sim \epsilon \omega$, the MSW resonance occurs, and 
the effective mixing angle becomes almost maximal, as it happens for 
the low energy mode shown in the figure. At $\lambda \gg \epsilon \omega$, 
the state $\nu_x$ decouples because of the suppression of the mixing angle 
in matter, making this an effectively two-flavor $\nu_e \leftrightarrow
\nu_y$ problem. The oscillations in ${\rm P}_8^{ey}$ have vanishing amplitude
and the motion in the triangle is restricted to the
$\nu_e$--$\nu_y$ edge.

\item
At even larger matter densities, $\lambda \gg \omega$,
the amplitude of $\nu_e \leftrightarrow \nu_y$
oscillations, which is the amplitude of ${\rm P}_3^{ey}$
oscillations, starts decreasing and the motion in the triangle
becomes more and more confined to be near the $\nu_e$
vertex as in the case of the high energy mode shown in the figure. 

\end{itemize}

All the above features may be understood analytically through
Eqs. (\ref{rhoeeP})-(\ref{3flavp8})
and the two-flavor evolution functions
\barr
\eta(\omega^{ey},\theta^{ey},0,r) & = &   
1- 2\sin^2 2\theta_{13} \sin^2 \left(\frac{h \omega r}{2}\right)\; ,\\
\eta(\omega^{ex},\theta^{ex},0,r) & = &   
1- 2\sin^2 2\theta_{12} \sin^2 \left(\frac{h\epsilon \omega r}{2}
\right)\;.
\earr
The above expressions are approximate, since we ignore the slowest 
frequency modes (depending on $\omega^{xy}$) and assume complete 
factorization. We find however, that these expressions agree 
with the numerical solution reasonably well .

In the case of finite but constant matter density, we use the angles 
$\theta^{\alpha\beta}$ and frequencies $\omega^{\alpha\beta}$ 
in matter, both of which are energy dependent. 
Note that the amplitudes in this case are proportional to 
$2 \sin^2 2\theta^{\alpha\beta}$ in matter and can be maximal 
(spanning a full edge of the triangle) 
when there is an MSW resonance.

When the matter density
encountered by the neutrino varies such that neutrinos pass
through an MSW resonance, they undergo flavor transitions with
adiabaticities depending on their energy, the relevant mixing
angle and the matter profile. In the limit of a small mixing
angle, a completely adiabatic $H$ resonance is represented by  
a reflection of the neutrino state about ${\bf e}_8^{ey}$ in
the ${\bf e}_3$--${\bf e}_8$ triangle.
A non-adiabatic $H$ resonance corresponds to a state that tries to 
move towards this reflected point, but does not completely succeed.
Passage through the $L$ resonance similarly corresponds to
a reflection about ${\bf e}_8^{ex}$.

\subsection{Synchronized oscillations}
\label{synchronized}

At extremely large neutrino densities, it is expected that
neutrinos of all energies oscillate synchronously with
a common frequency $\langle \omega ^{\alpha\beta}\rangle$ 
about $\Bvec^{\alpha\beta}$, 
given in the two-flavor case by \cite{wong-talk}
\beq
\langle \omega^{\alpha\beta} \rangle = \frac{\omega^{\alpha\beta}}{\omega}\;\frac{\Dvec\cdot\Dvec^{(1)}}{|\Dvec|^2} ~ ,
\label{omegasync}
\eeq
where ${\bf D}$'s are the moments defined in Eq. (\ref{moments-def}).
The frequency $\langle \omega^{\alpha\beta} \rangle$ 
crucially depends on the neutrino energy spectrum. 
The box-spectrum that we have chosen corresponds 
to $\la \omega^{ey} \ra \approx 0.49$ km$^{-1}$. 
In Fig.~\ref{fig:synchro} we show  
${\rm P}_3^{ey}, {\rm P}_8^{ey}$ and $\rho_{\nu_e \nu_e}$
as functions of the radial coordinate $r$ for synchronized 
neutrino oscillations.

\begin{figure}
\parbox{9cm}{
\epsfig{file=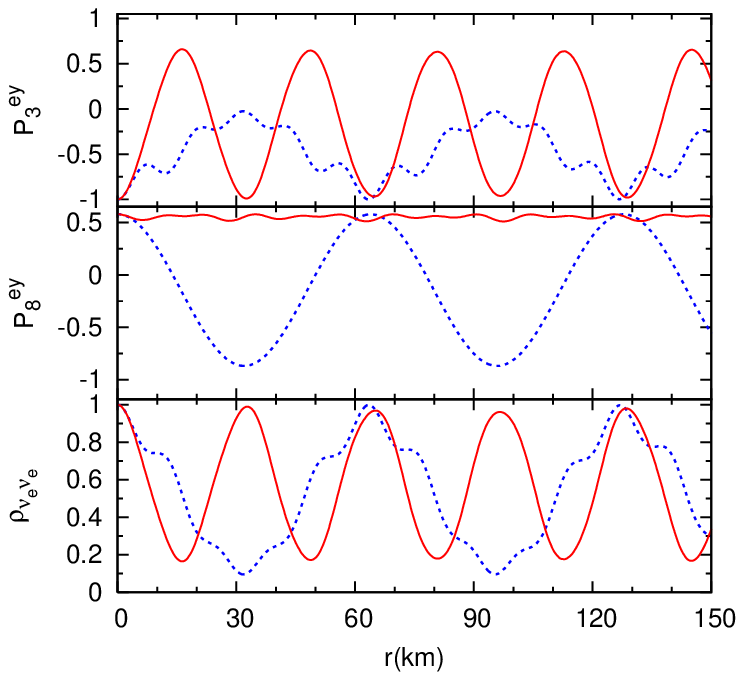,width=8.5cm}
}
\parbox{7cm}{
\epsfig{file=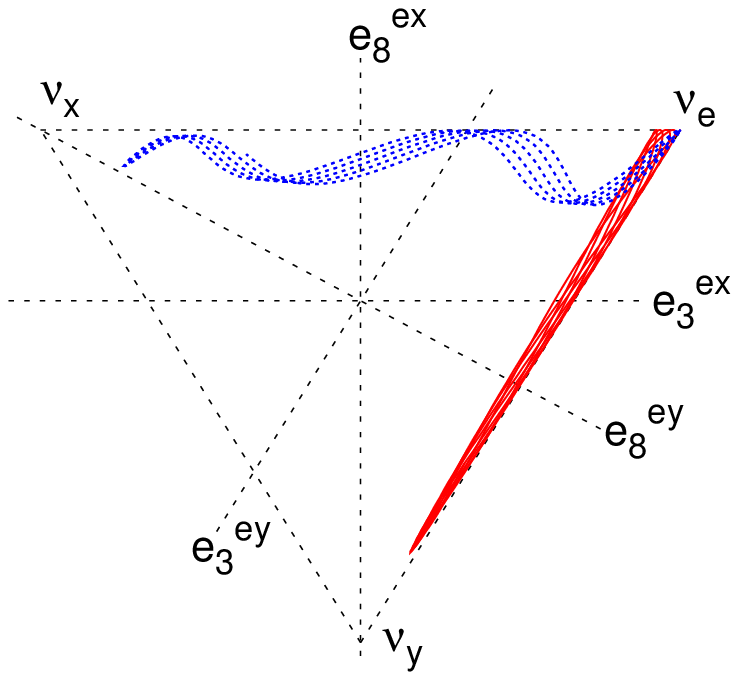,width=7cm}
}
\caption{Synchronized oscillations for neutrinos of
$E= 20.0$ and MeV $29.6$ MeV, which overlap completely.
We choose $\epsilon=1/5$ to emphasize the nature of oscillations, 
$\theta_{13}=0.2$ and  $\mu=100{~\rm km}^{-1}$.  
Oscillations in vacuum and matter are shown by
dotted (blue) and undotted (red) lines respectively. 
For matter, we choose normal hierarchy and $\lambda=0.5~{\rm km}^{-1}$.
 Note that the orbits on the triangle are the same for different energies.}
\label{fig:synchro}
\end{figure}

The following observations may be made from the figure:
\begin{itemize}

\item The observations in Sec.~\ref{vac-msw} remain true,
except that neutrinos of all energies oscillate with
a common frequency in vacuum in the two flavor limits of
each of the $\alpha - \beta$ subspaces. The response of
all neutrinos to the neutrino-neutrino potential is thus
identical.

\item Even in the presence of matter, the synchronized 
oscillation amplitude is independent of energy, unlike what 
happens for non-collective oscillations.

\item The amplitude of the slower oscillations is almost maximal
because, in the chosen example, $\lambda \sim \epsilon\la\omega\ra$.

\item The orbits drift periodically, even if $\omega$ and $\epsilon \omega$ 
are commensurate, because $\langle \omega \rangle$ and $\langle \epsilon \omega \rangle$ are not commensurate in general. This is due to corrections to Eq. (\ref{omegasync}) arising out of incomplete synchronization.

\end{itemize}

The above observations are explained analytically along the
same lines as the vacuum / MSW case. The two-flavor evolution
functions are given by
\barr
\eta(\omega^{ey},\theta^{ey},0,r) & = &   
1- 2\sin^2  \langle 2 \theta_{13} \rangle  
\sin^2 \left(\frac{h \langle \omega \rangle r}{2} \right)\; ,\\
\eta(\omega^{ex},\theta^{ex},0,r) & = &   
1- 2\sin^2  \langle 2 \theta_{12}\rangle  
\sin^2 \left(\frac{h\epsilon \langle \omega \rangle r}{2} \right)\;~. 
\earr
In the plots we see that fast oscillations have wavelength 
$2\pi/\omega\approx 12$ km. This matches the value of 
$\la \omega^{ey} \ra$ calculated from Eq. (\ref{omegasync}).

In the presence of a finite matter density, the MSW potential 
$\lambda$ also takes an effective average value given by 
\cite{wong-talk}
\beq
\langle \lambda \rangle = 
\lambda\;\frac{\Dvec \cdot \Svec}{|\Dvec|^2}~.
\eeq
Naturally, the mixing angle is also the same for all energies, since 
\beq
\sin^2 \langle 2 \theta^{\alpha\beta}\rangle= 
\frac{\sin^2 2\theta^{\alpha\beta}}{(\langle \lambda 
\rangle/\langle \omega^{\alpha\beta} \rangle-
\cos2\theta^{\alpha\beta})^2 +\sin^2 2\theta^{\alpha\beta}}~. 
\eeq
Thus not only the frequency, but also the amplitude of oscillations 
is universal in the synchronized limit. The MSW resonance is 
collective, occuring with the same adiabaticity for all 
neutrinos/antineutrinos at the same $\lambda$ when the 
relevant condition is met, as was shown in the two-flavor case 
\cite{wong-0203180,abazajian-beacom-bell-0203442,wong-talk}.
The factorization shown in Sec.~\ref{factorization} allows
the result to be extended to the three-flavor situation.

\subsection{Bipolar Oscillations}
\label{bipolar}

When the hierarchy is inverted and there are comparable numbers of neutrinos and antineutrinos
in the system, i.e. $\mu |\Dvec| \sim \omega |\Bvec|$, the 
influence of the $\omega$ and $\mu$ terms in the equations of motion 
depends crucially on the relative orientation of $\Dvec, \Bvec$ and the 
magnitude of $\Dvec$ itself. This subtle interplay gives rise to bipolar oscillations.

Many of the notions about bipolar oscillations in the two-flavor
formalism \cite{hannestad-raffelt-sigl-wong-0608095,duan-fuller-carlson-qian-0703776} remain valid with three flavors,
since they do not depend on the number of flavors, or equivalently,
on the dimensionality of the Bloch vectors.
The system is best understood in terms of the ``pendulum vector''
${\rm \bf Q}$ defined in the interaction frame as 
\cite{hannestad-raffelt-sigl-wong-0608095}
\beq
{\rm \bf Q} \equiv {\rm \bf S} - \frac{\omega}{\mu} \Bvec \; ,
\eeq
in terms of which the equations of motion are
\barr
\dot{\rm \bf Q} & = & \mu \Dvec \times {\rm \bf Q} - 
\frac{\omega}{\mu} \dot{\Bvec} \; , \\
\dot{\Dvec} & = & \omega \Bvec \times {\rm \bf Q} \; .
\earr
The antisymmetry of the generalized cross product in Eq. (\ref{cross-product})
implies that even in the
case of three flavors, $|{\rm \bf Q}|^2$ and $\Dvec \cdot \Bvec$
are conserved for large $\mu$.

In the two-flavor case, the motion can be understood in terms of
a spherical pendulum \cite{hannestad-raffelt-sigl-wong-0608095}, 
with the total energy
given by $\omega \Bvec \cdot {\rm \bf Q} + \mu |\Dvec|^2/2$.
For normal hierarchy, the pendulum is stable and executes only
small oscillations.
For inverted hierarchy, however, the system behaves like an inverted
pendulum, which tries to relax to its stable position.
The polarization vectors then remain almost static, but periodically
dip to the configuration with the lowest potential energy 
$\Bvec \cdot {\rm \bf Q}$.
Thus for inverted hierarchy, one can have a large flavor
swap during the dip.
The duration between successive dips is given by 
$\tau^{bip} \approx \sqrt{\omega \mu |{\rm \bf Q}|}$ with
logarithmic corrections depending on $\theta$ and $\lambda$. Since $\mu>\omega$, 
individual $\Pvec$ remain bound to each other, and therefore 
behave identically to $\rm \bf Q$.

Addition of a third flavor may  change the behaviour 
significantly, as we show in Figs.~\ref{fig:bipolar1}  and
\ref{fig:bipolar2} for two extreme values of $\lambda$.
We consider the case of inverted hierarchy, and a box-spectrum of energies 
$E=(1$--$51)$ MeV with the number of antineutrinos 
as $(1-\alpha)$ times the number of neutrinos, with $\alpha = 0.2$.
Given that the hierarchy in the solar sector is normal, we expect 
bipolar effect only in the $e-y$ subspace, combined with usual
neutrino oscillations in the $e-x$ subspace.
The following observations may be made from the figures:

\begin{figure}
\parbox{9cm}{
\epsfig{file=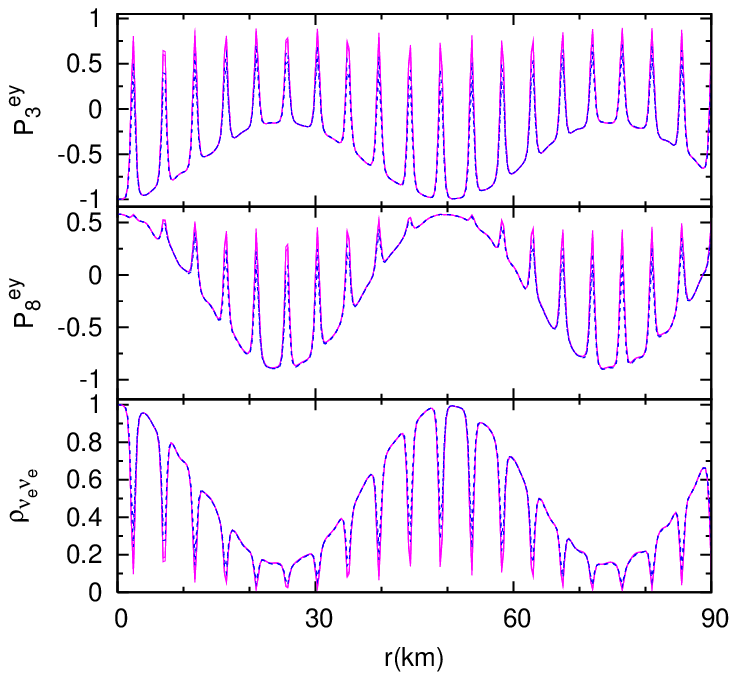,width=8.5cm}
}
\parbox{7cm}{
\epsfig{file=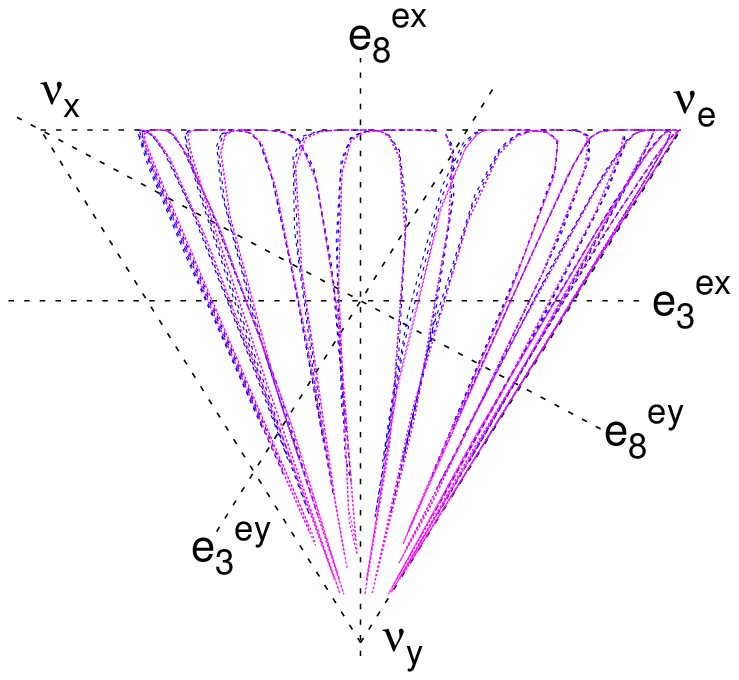,width=7cm}
}
\caption{Bipolar oscillations at small $\lambda$
for neutrinos (dotted, blue) and 
$20\%$ fewer antineutrinos (undotted, pink) of different energies, which almost overlap. 
We choose inverted hierarchy, $|\epsilon|=1/30$, $\theta_{13}=0.01, \mu=10{~\rm km}^{-1}$ and 
$\lambda=0.001~{\rm km}^{-1}$. Note that the plots are the same for different energies, because of strong collective behaviour.}
\label{fig:bipolar1}
\end{figure}

\begin{figure}
\parbox{9cm}{
\epsfig{file=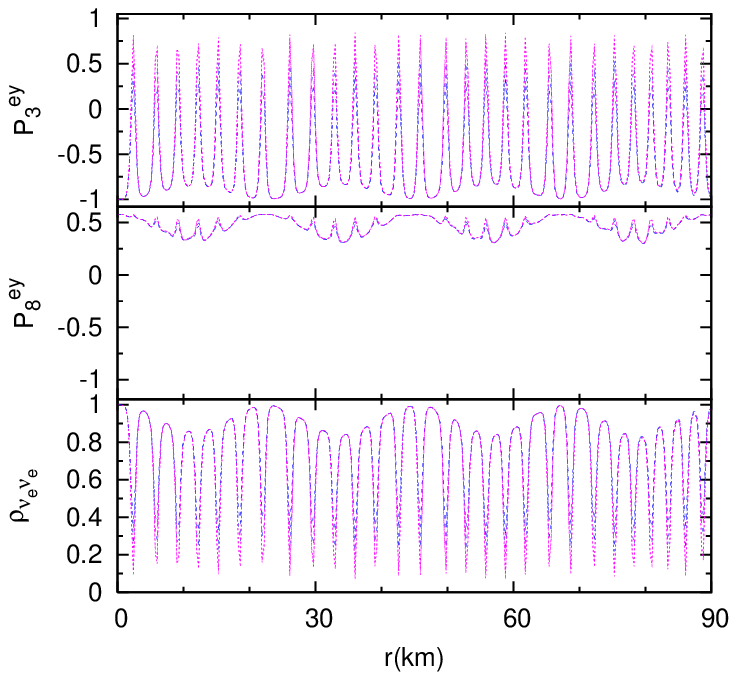,width=8.5cm}
}
\parbox{7cm}{
\epsfig{file=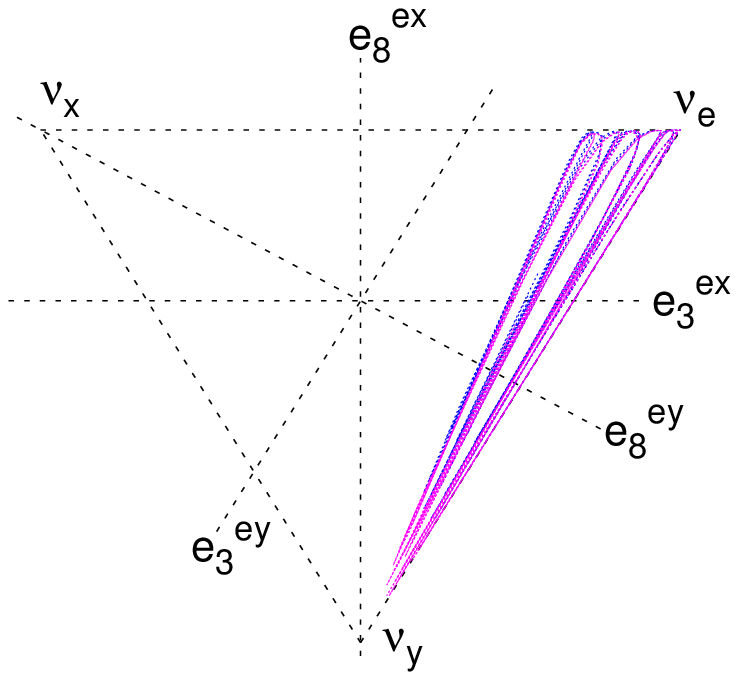,width=7cm}
}
\caption{Bipolar oscillations at large $\lambda$
for neutrinos (dotted, blue) and antineutrinos (undotted, pink) of different energies, which almost overlap. We choose inverted hierarchy, $|\epsilon|=1/30$, $\theta_{13}=0.01, \mu=10{~\rm km}^{-1}$ and $\lambda=0.3~{\rm km}^{-1}$. Note that the plots are the same for different energies, because of strong collective behaviour.}
\label{fig:bipolar2}
\end{figure}

\begin{itemize}

\item The evolution of both ${\rm P}_3^{ey}$ and ${\rm P}_8^{ey}$ 
consists of a series of bipolar ``kinks'' as in the two-flavor
case \cite{hannestad-raffelt-sigl-wong-0608095}, 
modulated by an envelope with the
frequency $\la \omega^{ex} \ra$. The evolutions for neutrinos and
antineutrinos closely follow one another, which is expected
from the conservation of $\Bvec \cdot \Dvec$.

\item Significant three-flavor effects are present for small
$\lambda$, since the whole triangle is seen to be filled with 
oscillations, forming a ``petal structure'' (Fig.~\ref{fig:bipolar1}).
It may be interpreted as a combination of slow
$\nu_e \leftrightarrow \nu_x$ oscillations and bipolar
oscillations that tend to take the state towards $\nu_y$ 
in periodic bursts.

\item The extent of motion towards $\nu_y$ depends mainly on
the asymmetry $\alpha$, whereas that towards $\nu_x$ depends
on $\sin^2 2\theta_{12}$.

\item For large $\lambda$ (Fig.~\ref{fig:bipolar2}), the oscillations 
in the $e-x$ sector are suppressed since the effective mixing angle 
$\theta_{12}$ in matter becomes small. 
The amplitude of the bipolar motion is however not affected substantially.

\end{itemize}

Bipolar oscillations (even in the two-flavor limit) 
do not have a sinusoidal form,
hence they are not associated with a fixed frequency.
They may be looked upon as a combination of a low
frequency (during the time that the $\nu_y$ component
is stationary, which we shall call the A phase) 
and a high frequency (the sudden dip
towards $\nu_y$, which we shall call the B phase). 
Therefore, our prescription in Sec.~\ref{three-flavor}
has to be applied with care. 
Note that the order of evolution matrices in Eq. (\ref{s-matrices})
is supposed to be in the decreasing order of frequencies.
Even if we neglect the slow evolution due to $\Bvec^{xy}$,
strictly speaking during the A phase, one should use 
the order ${\textsf S}^{ex} {\textsf S}^{ey}$ and during 
the B phase, the order should be 
${\textsf S}^{ey} {\textsf S}^{ex}$.
However, we find numerically that the evolution
${\textsf S}^{ex} {\textsf S}^{ey}$ closely matches the 
three-flavor solution over the complete evolution.
This therefore may be taken to be the heuristic solution 
for the bipolar oscillations in the three-flavor case.

We have not considered normal hierarchy, in which we expect
that starting with $\nu_e$  we'll have a stable system that
will not undergo bipolar oscillations, whereas starting with
$\nu_x$ or $\nu_y$, we'll have independent bipolar oscillations
towards $\nu_e$. It will be interesting to analyze the details 
of such a scenario, however it is beyond the scope of this paper.

\subsection{Spectral splitting}
\label{split}

As a system of neutrinos and antineutrinos transits from the 
collective regime ($\mu \gg \omega$) to vacuum ($\mu \sim 0$), the 
polarization vectors $\Pvec$ keep trying to align 
with ${\rm \bf H}$ in the adiabatic approximation. 
Due to the conservation of $\Bvec \cdot \Dvec$, as shown
in Sec.~\ref{bipolar}, this alignment is not possible 
for all $\Pvec$. 
Indeed, neutrinos with high energy need to flip over 
and anti-align with $\Bvec$ (which equals ${\rm \bf H}$ 
in vacuum) \cite{raffelt-smirnov-0705.1830,raffelt-smirnov-0709.4641}.
This leads to a sharp split in the energy spectrum, 
with the high energy $\nu_e$ getting completely
converted to the non-electron flavor and vice versa. 

A crucial requirement for the splits to develop is 
the preparation of the sytem for the split by 
the generation of components of $\Pvec$
that are transverse to $\Bvec$.
Bipolar oscillations do this easily for inverted hierarchy,
independent of matter effects. 
For normal hierarchy, in the presence of large matter effects 
the oscillations are suppressed, but an MSW resonance can prepare 
the transverse components. 

For illustration, we choose two situations, with large
and small $\lambda$ (Fig.~\ref{fig:split1} and \ref{fig:split2}
respectively) and the hierarchy is taken to be inverted. We choose the box-spectrum for $\nue$ 
and $\nuebar$ energies and the flux asymmetry $\alpha=0.33$.
We observe the following from the figures:

\begin{figure}
\parbox{9cm}{
\epsfig{file=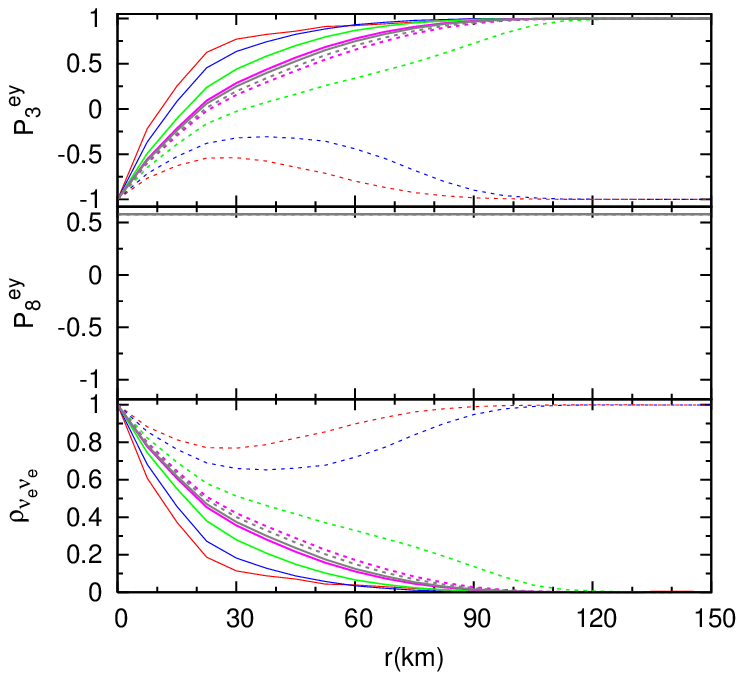,width=8.5cm}
}
\parbox{7cm}{
\epsfig{file=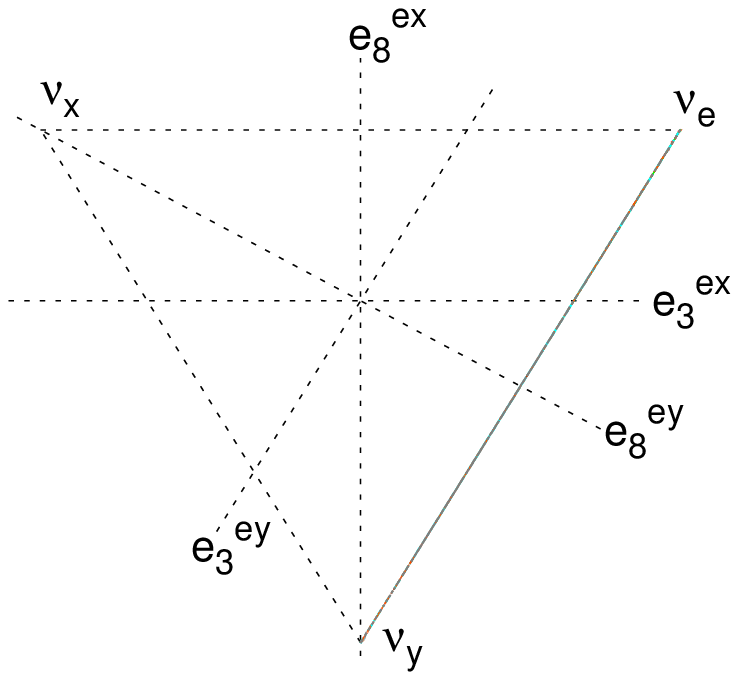,width=7cm}
}
\caption{Spectral splits at large $\lambda$ 
for neutrinos (dotted) and $33\%$ fewer 
antineutrinos (undotted) with a box-spectrum of energies 
$E=(1$--$51)$ MeV. 
In the $\rho_{\nu_e \nu_e}$ plot, the energy of neutrinos
(antineutrinos) increases (decreases) top downwards.
The energies (in MeV) of the modes, shown in the figure, 
are $1.0$ (Red), $1.5$ (Blue), $3.5$ (Green), $12.5$ (Pink) 
and $32.0$ (Grey). 
We take inverted hierarchy, $|\epsilon|=1/30$, 
$\theta_{13}=0.01$, $\mu=10^5~(50/r({\rm km}))^4{~\rm km}^{-1}$ and $\lambda=10~{\rm km}^{-1}$.
In the ${\bf e}_3$--${\bf e}_8$ triangle, the evolution is always
along the $\nu_e$--$\nu_y$ edge.
}
\label{fig:split1}
\end{figure}

\begin{figure}
\parbox{9cm}{
\epsfig{file=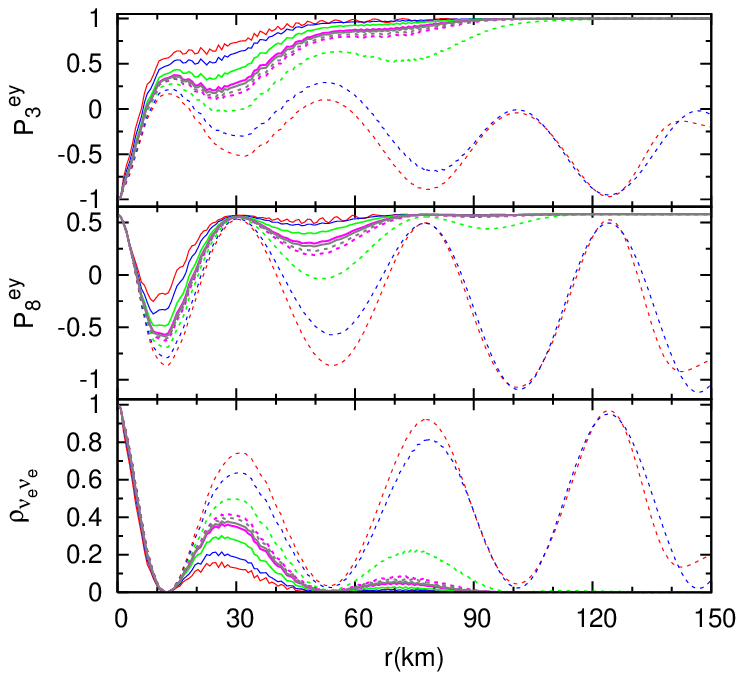,width=8.5cm}
}
\parbox{7cm}{
\epsfig{file=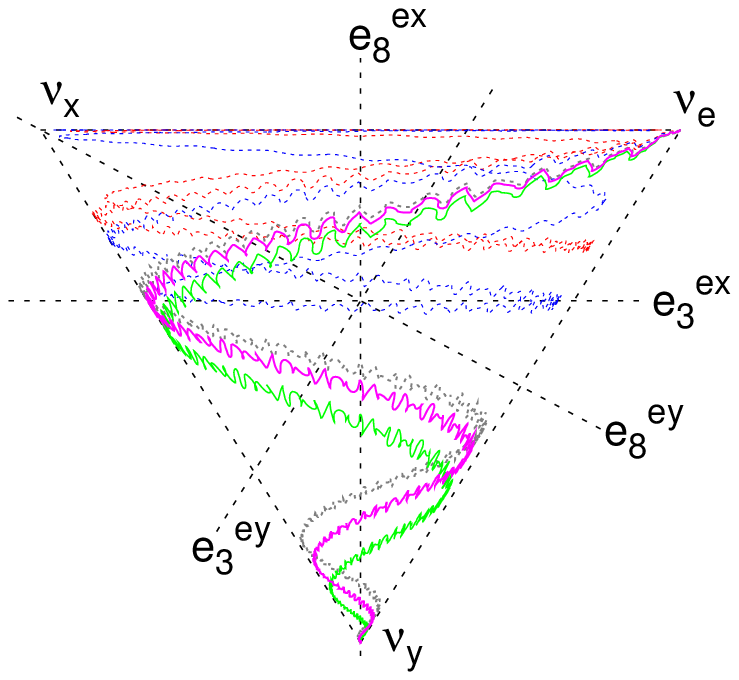,width=7cm}
}
\caption{Spectral splits at small $\lambda$ 
for neutrinos (dotted) and $33\%$ fewer 
antineutrinos (undotted) with a box-spectrum of energies 
$(E=1$--$51)$ MeV. 
The conventions for lines is the same as that in 
Fig.~\ref{fig:split1}.
We take inverted hierarchy, $\epsilon=1/30$, $\theta_{13}=0.01$ $\mu=10^5~(50/r({\rm km}))^4{~\rm km}^{-1}$ and $\lambda=0.1~{\rm km}^{-1}$.
In the ${\bf e}_3$--${\bf e}_8$ triangle, we show only some of the
representative energies that have different behaviours.
}
\label{fig:split2}
\end{figure}

\begin{itemize}
\item For large $\lambda$, there is only a single split for neutrinos,
which can be seen in ${\rm P}_3^{ey}$.  
The split is not visible in the triangle since the neutrinos
are confined to the $\nu_e$--$\nu_y$ edge.
However, the low energy neutrinos move towards $\nu_e$
and the high energy ones towards $\nu_y$.

\item for small $\lambda$, the split is not only in 
${\rm P}_3^{ey}$ but also in ${\rm P}_8^{ey}$. 
There also are oscillations with large amplitudes.
Some neutrino states drift towards and ultimately 
reach $\nu_y$, 
while the others keep oscillating between $\nu_e$ and $\nu_x$.

\end{itemize}

The above observations can be understood as follows.
For large $\lambda$, the solar mixing angle is suppressed 
and hence the problem reduces essentially to a two-flavor one
in the $e-y$ subspace.
Thus, the split is only in ${\rm P}_3^{ey}$.
The split happens in neutrinos since there are more neutrinos
than antineutrinos at any given energy.
For small $\lambda$, in addition to the above split,
there are large $\nu_e \leftrightarrow \nu_x$ oscillations,
which give rise to a split that is observable also
in ${\rm P}_8^{ey}$, which was absent for large $\lambda$.

A detailed understanding of the spectral splits in the
three-flavor case, including predictions for the positions
of the spectral splits has been obtained \cite{future-split}.

\section{Collective effects on neutrinos in SN}                  
\label{full-sn}

Collective effects are likely to be important in the context of 
neutrinos emitted from a SN. The number density of neutrinos and 
antineutrinos streaming off the neutrinosphere is quite large, 
so $\mu$ dominates over $\lambda$ and $\omega$ 
upto a radius of a few ten or hundred kilometers respectively.  
Therefore, it is likely that one or more flavor conversion mechanisms identified in 
Sec.~\ref{toy-cases} come into play in the different regions
inside the star.

In this section, we study the effect of collective oscillations
and their interplay with subsequent MSW transitions inside a SN. 
To illustrate the nature of these effects, we numerically solve for the 
flavor evolution equations by taking a realistic SN density 
profile, including the collective effects. We present results primarily  
for inverted hierarchy, because collective effects are not expected to play a 
significant part for normal hierarchy \footnote{However, during the neutronization 
burst phase of an O-Ne-Mg SN, this need not be true.}. In the numerical study, we take 
$|\dmsq_{13}|=2.5\times10^{-3}$ eV, $|\epsilon|=1/30$, $\theta_{12}=0.6$, and two representative values of $\theta_{13}$, viz.
$\theta_{13}=0.001$ and $0.1$. 
We then apply the formalism developed in 
Sec.~\ref{formalism} to the case of neutrinos 
streaming from the SN neutrinosphere and identify the regimes
where different flavor conversion mechanisms are at work.
This allows us to explain the features in the observable 
neutrino and antineutrino spectra, and understand the
three-flavor effects.

\subsection{SN model and parameters for numerical simulation}
\label{sn-numbers}
The SN model is defined by the following choice for the emission geometry, 
initial flavor dependent spectra and fluxes, the collective potential and 
the matter density profile. We would like to emphasize that these choices are 
canonical and more importantly, the specific value of the luminosity or the 
spatial dependence of the collective potential does not affect results significantly.
Any large initial value of $\mu$ (such that it exceeds $\omega$) and its slow decrease with $r$ gives almost identical results. In other words, the results are not fine-tuned. However, the results would  depend on the flavor spectra and the matter density profile,
as these determine the initial conditions, the collective potential
and the effective mixing parameters.

\subsubsection{Emission geometry}  

Neutrinos with different energies and flavors start
freestreaming at different $r$, but flavor evolution 
does not start until much later. Thus the radius of the neutrinosphere $r_0$ 
is used only to set the initial conditions. We therefore use the 
``bulb-model'' of neutrino emission from the SN as discussed in \cite{duan-fuller-carlson-qian-0606616} 
with a nominal neutrinosphere at $r_0=10\;{\rm km}$. We assume steady-state 
half-isotropic emission from the neutrinosphere.
\subsubsection{Initial spectra and fluxes}
The flavor-dependent primary neutrino spectra at $r_0$ are parametrized as \cite{keil}
\beq
F_{\nu_\alpha}(E)=
\Phi_{\nu_\alpha} \frac{N(\xi_\alpha)}{\la {E}_{\nu_{\alpha}}\ra}
\left(\frac{E}{\la {E}_{\nu_{\alpha}}\ra}\right)
^{\xi_\alpha}{\rm exp}\left[-(\xi_\alpha+1)\frac{E}{
\la {E}_{\nu_{\alpha}}\ra}\right]~,
\eeq
where $N(\xi)=(1+\xi)^{1+\xi}/\Gamma(1+\xi)$. This spectrum is 
normalized such that 
$\int_0^{\infty} dE_{\nu_{\alpha}}
~F_{\nu_{\alpha}}(E_{\nu_{\alpha}})
= \Phi_{\nu_\alpha}$ and has the average energy 
$\la {E}_{\nu_{\alpha}}\ra$. 
The above parametrization has the advantage that the spectra can be 
analytically integrated, including the effects of spectral pinching through $\xi_{\alpha}$.
The number flux $\Phi_{\nu_{\alpha}}$ is given by
$\Phi_{\nu_{\alpha}}=L_{\nu_\alpha}/\la {E}_{\nu_{\alpha}}\ra$, where 
$L_{\nu_\alpha}$ is the luminosity in the flavor $\nu_\alpha$. We remark that an equipartition in the luminosity is often assumed for simplicity. For illustration, we choose the above parameters as 
\barr
{L_{\nu_{\alpha}}}&=& 1.5\times10^{51}{\rm ergs/sec}\;, \quad \xi_\alpha=3 \; \, \nonumber\\
{\la {E}_{\nue} \ra}&=&10\;{\rm MeV} \; , \quad
{\la {E}_{\nuebar}\ra}=15\;{\rm MeV} \; , \quad
{\la {E}_{\nu_{x,y},\;\nubar_{x,y}}\ra }=20\;{\rm MeV} \;.
\label{SNparameters}
\earr

Remembering that $E\sim p=|\Delta m^2_{13}|/(2\omega)$, 
we can rewrite the above information in terms of $\omega$,
if desired. 
Combining Eq. (\ref{pvecinitial}) with the definitions of moments
in (\ref{moments-def}), allows us to calculate the 
values of $\Dvec(r_0),\;\Svec(r_0)$ and $\Dvec^{(1)}(r_0)$
for the above spectrum as  
\barr
\Dvec(r_0)&=&\frac{\left(\la E_{\nuebar} \ra - \la E_{\nue} \ra\right)\la 
E_{\nux} \ra}{\la E_{\nue} \ra\la E_{\nux} \ra + \la E_{\nuebar} \ra 
\left(4\la E_{\nue} \ra+\la E_{\nux} \ra\right)}\;{\bf e}_{e}= 
\frac{1}{11}\;{\bf e}_{e}\;,\\
\Svec(r_0)&=&\frac{\left(\la E_{\nue} \ra+\la E_{\nuebar} \ra\right)\la 
E_{\nux} \ra - 2\la E_{\nue} \ra\la E_{\nuebar} \ra}{\left(\la E_{\nue} \ra
+\la E_{\nuebar} \ra\right)\la E_{\nux} \ra + 4\la E_{\nue} \ra\la 
E_{\nuebar} \ra}\;{\bf e}_{e}= \frac{2}{11}\;{\bf e}_{e}\;,\\
\Dvec^{(1)}(r_0)&=&\frac{2\dmsq_{13}}{3}\;\frac{1/\la E_{\nue} \ra^2 
+ 1/\la E_{\nuebar} \ra^2 - 2/\la E_{\nux} \ra^2}{1/\la E_{\nue} \ra + 
1/\la E_{\nuebar} \ra + 4/\la E_{\nux} \ra}\;{\bf e}_{e} = 0.215\; 
{\bf e}_{e}\; {\rm km}^{-1}\;.
\label{DSD1exprn}
\earr
Using the above expressions, 
$\la \omega \ra \equiv \Dvec\cdot\Dvec^{(1)}/|\Dvec|^2$ is calculated to be 
\beq
\la \omega \ra = 2.37 ~{\rm km}^{-1} \; ,
\label{omega-value}
\eeq
which allows us to write $\la \omega^{ey} \ra = \la \omega \ra$ and
$\la \omega^{ex} \ra = \epsilon \la \omega \ra$ in terms of $\la \omega\ra$, as per Eq. (\ref{omegasync}).

\subsubsection{Collective potential and matter density profile}

The collective potential for $r>r_0$ for the choice of parameters 
in Eq.~(\ref{SNparameters}) is given by 
\beq
\mu(r)=0.45\times10^{5} \; g(r)\;{\rm km^{-1}}\;,
\eeq
where $g(r)$ is given in Eq. (\ref{dilfac}).
For illustration, we choose the shock-wave simulation inspired
density profile that corresponds to 
\footnote{This is the same as the one used in
\cite{fogli-lisi-mirizzi-montanino-0304056} at t=4 sec.}
\beq
\lambda(r)=1.84\times10^6/r^{2.4}\;{\rm km^{-1}}\;.
\eeq
The profiles of $\lambda(r)$ and $\mu(r)$ are shown in
Fig.~\ref{wlm}.

\begin{figure}
\epsfig{file=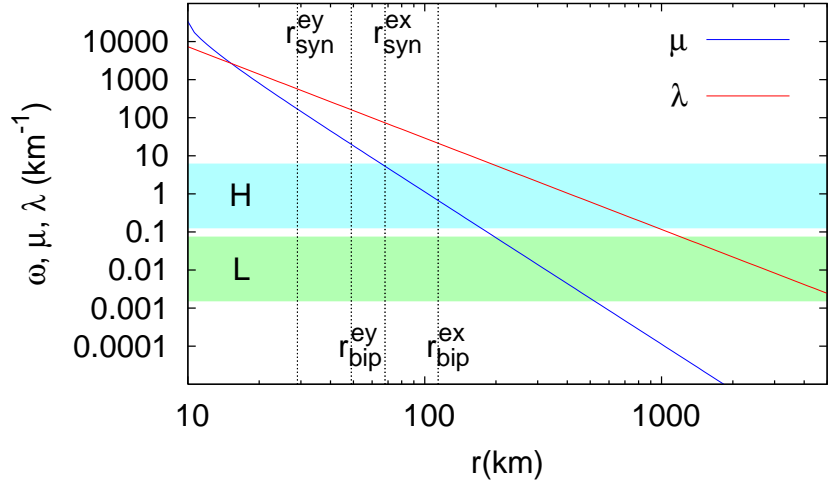,width=12cm}
\caption{The profiles of $\lambda(r)$ and $\mu(r)$ for 
the SN model chosen in this section, along with the
bands for the MSW resonances $H$ and $L$.
Also indicated are the terminal values of $r$ where synchronized /bipolar oscillations 
for the $e-y$ and $e-x$ flavors take place.
}
\label{wlm}
\end{figure}

\subsection{Flavor conversions inside a supernova}
\label{full-probs}

In this section we solve for the evolution of the neutrino
density matrix numerically for the chosen density profile, 
and show the neutrino flavor conversions. 
We expect synchronized oscillations in the region where
$\mu > 4 \la \omega^{ey} \ra \, {\rm S}_3^{ey} \,/\, 
({\rm D}_3^{ey})^2 \approx 208$ km$^{-1}$
\cite{fogli-lisi-marrone-mirizzi-0707.1998}, which corresponds to 
$r_{syn}^{ey} \approx 30$ km in our example.
In the further region till $\mu \approx \la \omega^{ey} \ra \, / \,
{\rm D}_3^{ey} \approx 26$ km$^{-1}$ 
\cite{fogli-lisi-marrone-mirizzi-0707.1998},
which corresponds to $r_{bip}^{ey} \approx 49$ km, 
$\nu_e \leftrightarrow \nu_y$ bipolar oscillations
are expected.
Beyond this region
the spectral split in the $e-y$ sector should develop,
and subsequently MSW resonances should take place.
Similarly we calculate for the $e-x$ flavors, the relevant 
values of $r_{syn}^{ex}\sim 68$ km and 
$r_{bip}^{ex}\sim 114$ km for approximate boundaries of 
synchronized and 
bipolar oscillations in the $e-x$ sector. 
However, no bipolar oscillations take place
in the $e-x$ sector since the corresponding hierarchy is normal.
In Fig.~\ref{wlm}, we show the positions corresponding to
$r_{syn}^{ey}, r_{bip}^{ey}, r_{syn}^{ex}$ and 
$r_{bip}^{ex}$.

\subsubsection{Small $\theta_{13}$}

Fig.~\ref{fig:smallt13} shows the flavor evolutions in terms of
${\rm P}_3^{ey}, {\rm P}_8^{ey}$ and the ${\bf e}_3$--${\bf e}_8$
triangle for neutrinos as well as antineutrinos, for
$\theta_{13} = 0.001$. This small value of $\theta_{13}$
ensures that the MSW resonance $H$ in antineutrinos is
nonadiabatic, so that the effects of this resonance are
not felt. One can then cleanly identify the collective
effects.
We make the following observations and interpretations 
based on the figure:

\begin{figure}
\parbox{9cm}{
\epsfig{file=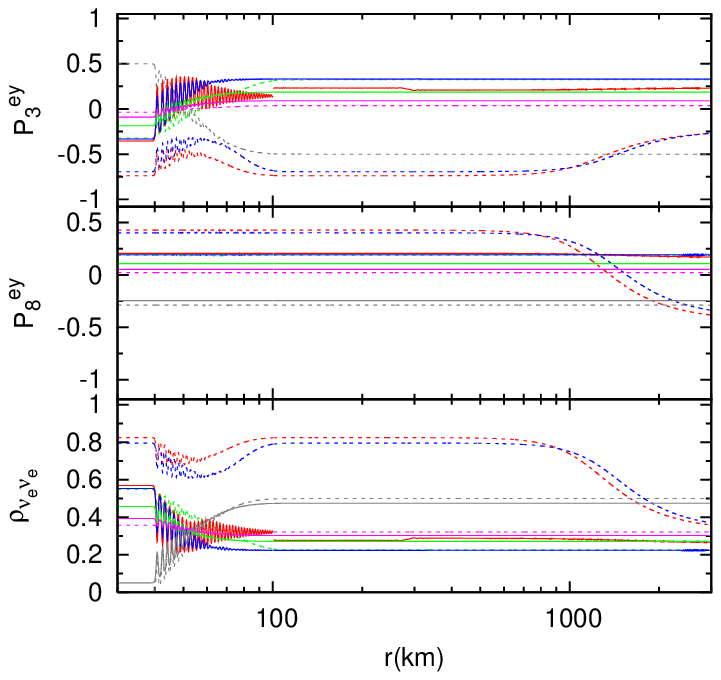,width=8.5cm}
}
\parbox{7cm}{
\epsfig{file=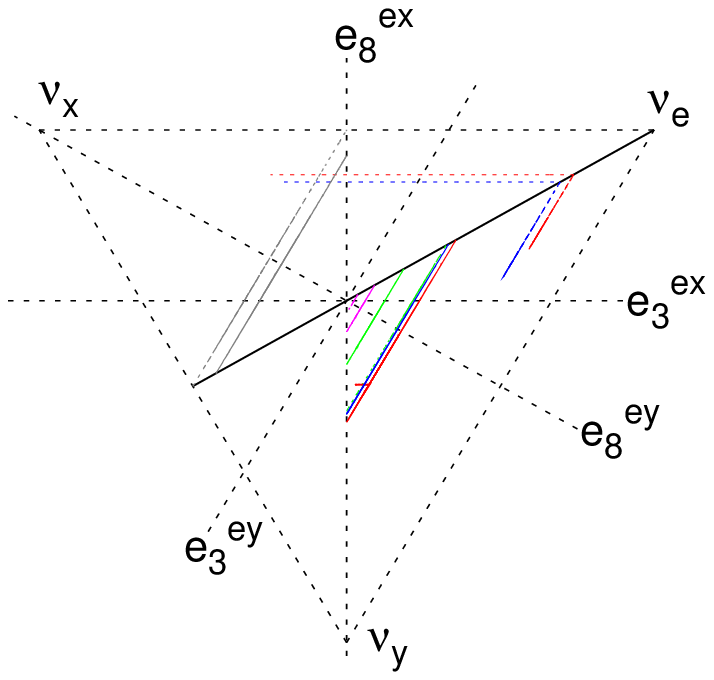,width=7cm}
}
\caption{The flavor evolution of representative energy modes of $\nu_e$(dotted) 
and $\nuebar$ (undotted) for the density profile in
Fig.~\ref{wlm}, with $\theta_{13}=0.001$. 
In the $\rho_{\nu_e \nu_e}$ plot, the energy of neutrinos as
well as antineutrinos increases top downwards.
The energies (in MeV) of the modes, shown in the figure, are $2.5$ (Red), $3.6$ (Blue), $9.4$ (Green), $13.3$ (Pink) and $50.0$ (Grey).
In the triangle plot, the bold line passing through $\nu_e$ 
is where all the 
neutrino and antineutrino states initially lie.
}
\label{fig:smallt13}
\end{figure}

\begin{itemize}

\item All the neutrinos and antineutrinos initially lie
on a line passing through $\nu_e$ in the ${\bf e}_3$--${\bf e}_8$
triangle. This is because the initial conditions are
taken to be symmetric in $\nu_x$ and $\nu_y$.

\item The flavor evolution starts only at $r = 40$ km,
which is slightly beyond $r_{syn}^{ey}$.
Before this point, the oscillations are synchronized, with
a vanishing amplitude since $\lambda \gg \la \omega^{ey} \ra$.

\item Between $r= 40$ and $60$ km, $\nu_e \leftrightarrow \nu_y$
bipolar oscillations are observed as rapid dips in ${\rm P}_3^{ey}$,
and consequently in $\rho_{\nu_e \nu_e}$.
These oscillations vanish when $r \gsim r_{bip}^{ey}$.

\item Around $r \approx 60$ km, a spectral split develops in
neutrinos along ${\rm P}_3^{ey}$. 
The spectral split tends to keep the low energy neutrinos at 
their original position, while taking the high energy neutrinos 
as well as almost all antineutrinos towards 
${\rm P}_3^{ex}=0$. \footnote{
There seems to be a spectral split in antineutrinos as well, at very low energies.
This is similar to the observation in 
\cite{fogli-lisi-marrone-mirizzi-0707.1998},
and may be the effect of
nonadiabaticity in the splitting.}

\item Between $r \approx 100-1000$ km, antineutrinos of different
energies undergo the $H$ resonance. 
However the resonance is highly nonadiabatic and does
not cause any flavor conversion. 

\item At $r=1000$ km and beyond, the effects of the MSW resonance
$L$ come into play, resulting in $\nu_e \leftrightarrow \nu_x$
conversion.
Since the high energy neutrinos are already close to
${\rm P}_3^{ex} =0$, there is effectively no flavor conversion.
However the low energy neutrinos tend to convert to
$\nu_x$, which is seen as
a movement parallel to the $\nu_e$--$\nu_x$ edge
in the ${\bf e}_3$--${\bf e}_8$ triangle.

\item Since all the flavor conversions can be understood as
a net effect of two-flavor phenomena taking place in 
well-separated regions in the star, the flavor transitions
in the ${\bf e}_3$--${\bf e}_8$ triangle are always
parallel to the $\nu_e$--$\nu_x$ edge or $\nu_e$--$\nu_y$
edge.

\end{itemize}

Thus, for a small $\theta_{13}$, the collective effects can be 
clearly identified, whereas the effects due to the $H$ resonance are absent.
We calculate the flavor evolution till $r=5000$ km.
The collective effects have almost vanished by this time.
Further MSW resonances due to the shock wave
\cite{schirato-fuller-0205390,takahashi-sato-dalhed-wilson-0212195,fogli-lisi-mirizzi-montanino-0304056,tomas-kachelreiss-raffelt-dighe-janka-scheck-0407132,fogli-lisi-mirizzi-montanino-0412046,huber,dasgupta-dighe-0510219},
as well as possible effects of stochastic density
fluctuations or turbulence 
\cite{fogli-lisi-mirizzi-montanino-0603033,choubey-harries-ross-0605255,friedland-gruzinov-0607244}
will govern flavor
conversions here onwards.
Our calculations thus provide initial conditions for
neutrino spectra at this point.

\begin{figure}
\parbox{8cm}{
\epsfig{file=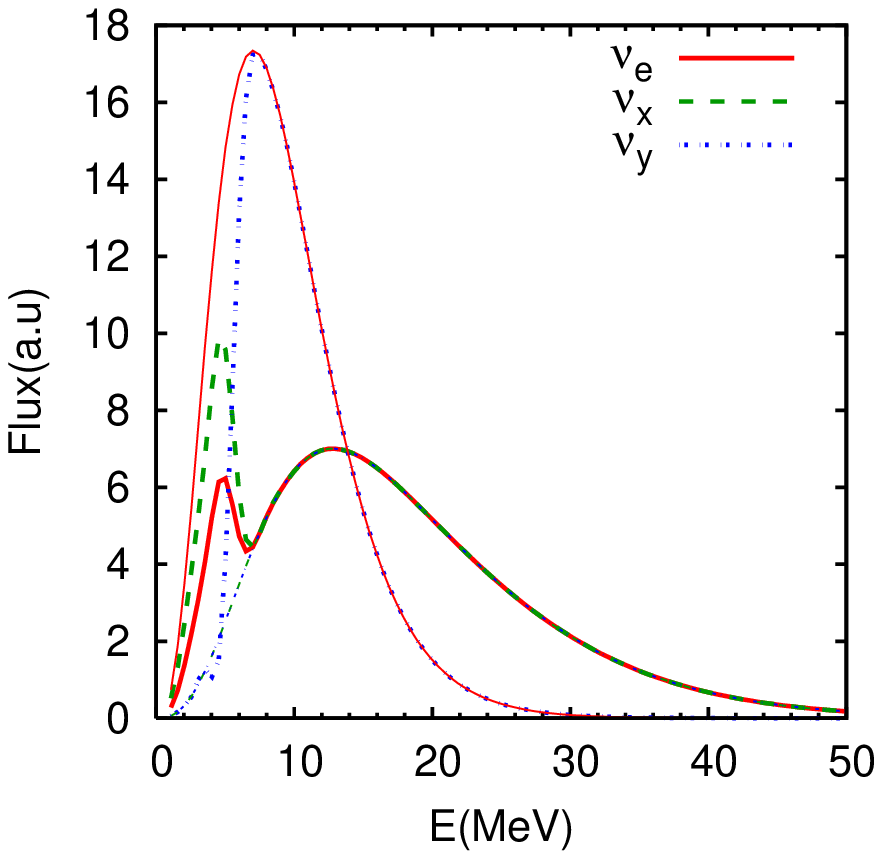,width=7cm}
}
\parbox{8cm}{
\epsfig{file=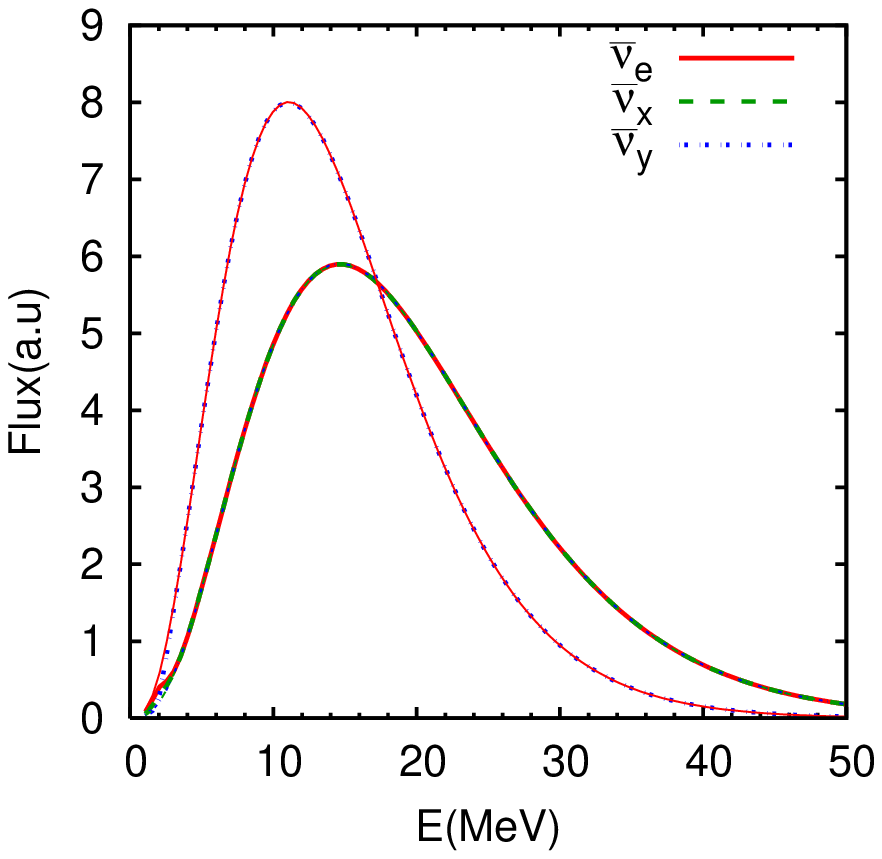,width=7cm}
}
\caption{Neutrino and antineutrino spectra at $r=5000$ km 
for $\theta_{13}=0.001$. The $e,\;x$ and $y$ flavors are shown in red(solid), green(dashes) and blue(dots). The thin lines/dashes/dots are for initial spectra and thick ones for the final spectra. The $\nu_e$ and $\nu_y$ spectra get swapped for  $E \gsim 7$ MeV,
whereas the lower energy $\nu_e$ spectrum partially mixes with
$\nu_x$. In the antineutrino sector, the $\nuebar$ and $\nubar_y$ spectra are almost completely swapped, while the $\bar{\nu}_x$ spectrum remains unaffected.}
\label{spectra-smallt13}
\end{figure}

In Fig.~\ref{spectra-smallt13}, we show the neutrino and
antineutrino spectra at $r=5000$ km.
We see that $\nu_e$ with $E \gsim 7$ MeV convert almost
completely to $\nu_y$ due to the spectral split,
whereas lower energy $\nu_e$ convert partially to
$\nu_x$ at the $L$ resonance.
In the antineutrino sector, the $\nuebar$ and $\nubar_y$ spectra are almost all completely 
swapped due to the spectral split, while
the $\bar{\nu}_x$ spectrum remains unaffected.

\begin{figure}
\parbox{9cm}{
\epsfig{file=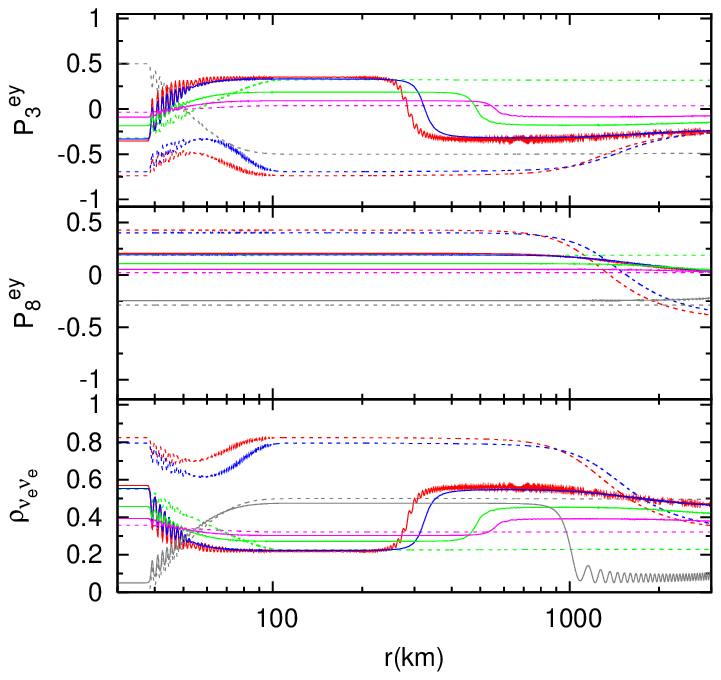,width=8.5cm}
}
\parbox{7cm}{
\epsfig{file=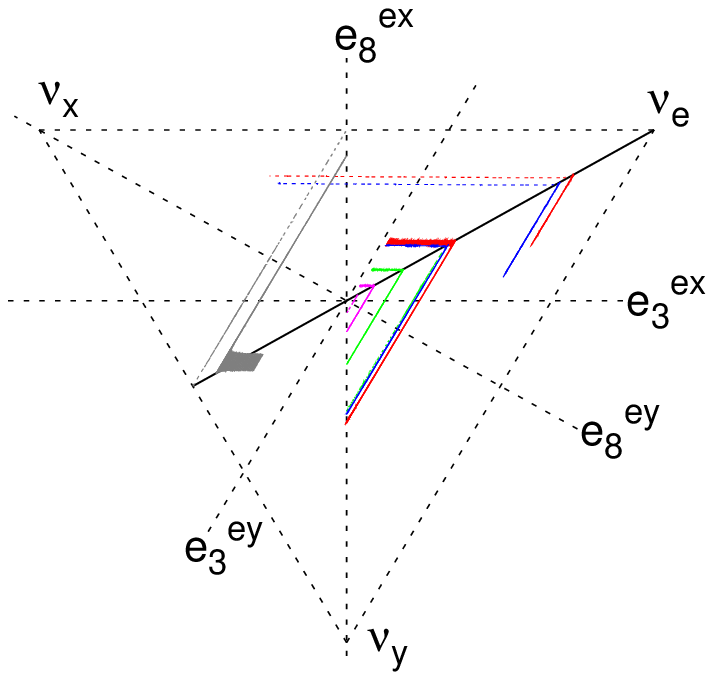,width=7cm}
}
\caption{The flavor evolution of same representative energy modes of $\nu_e$(dotted) 
and $\nuebar$ (undotted) for the density profile in
Fig.~\ref{wlm}, with $\theta_{13}=0.1$. 
The convention for the lines is the same as in Fig.~\ref{fig:smallt13}.
}
\label{fig:larget13}
\end{figure}

\subsubsection{Large $\theta_{13}$}

At large $\theta_{13}$ values, the $H$ resonance 
at $r \approx 100$--$1000$ km is adiabatic, and 
causes significant flavor conversions in antineutrinos.
In Fig.~\ref{fig:larget13}, we show the 
flavor evolution for $\theta_{13} = 0.1$. 
While the signatures of synchronized and bipolar oscillations
as well as the spectral split remain identical to the
$\theta_{13} = 0.001$ case, the $H$ resonance can be
seen to change the antineutrino picture substantially.
The conversions in the neutrino sector, on the other hand, 
are identical to the small $\theta_{13}$ case. The following 
observations can be made from the figure.

\begin{itemize}
\item The spectral split gives rise to a complete 
$\nuebar$--$\bar{\nu}_y$ conversion,
which takes antineutrinos to ${\rm P}_3^{ex}=0$.

\item The $H$ resonance again swaps the $\nuebar$--$\nubar_y$
spectra, thus undoing the earlier effect of the spectral split.
This takes the antineutrinos back to their starting position
in the triangle.

\item Antineutrinos are now not on the ${\rm P}_3^{ex}=0$ line
as in the small $\theta_{13}$ case. As a result, the large
value of $\theta_{12}$ causes substantial $\nuebar$--$\nubar_x$
conversion as the neutrinos emerge from the $L$ resonance
region.

\end{itemize}

\begin{figure}
\parbox{8cm}{
\epsfig{file=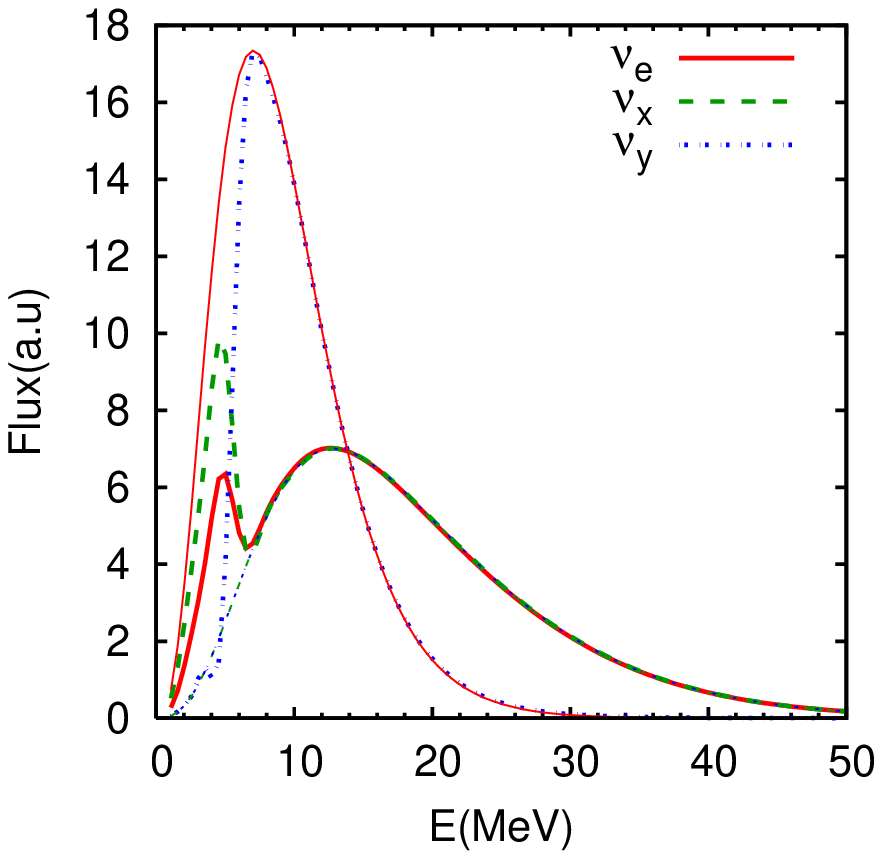,width=7cm}
}
\parbox{8cm}{
\epsfig{file=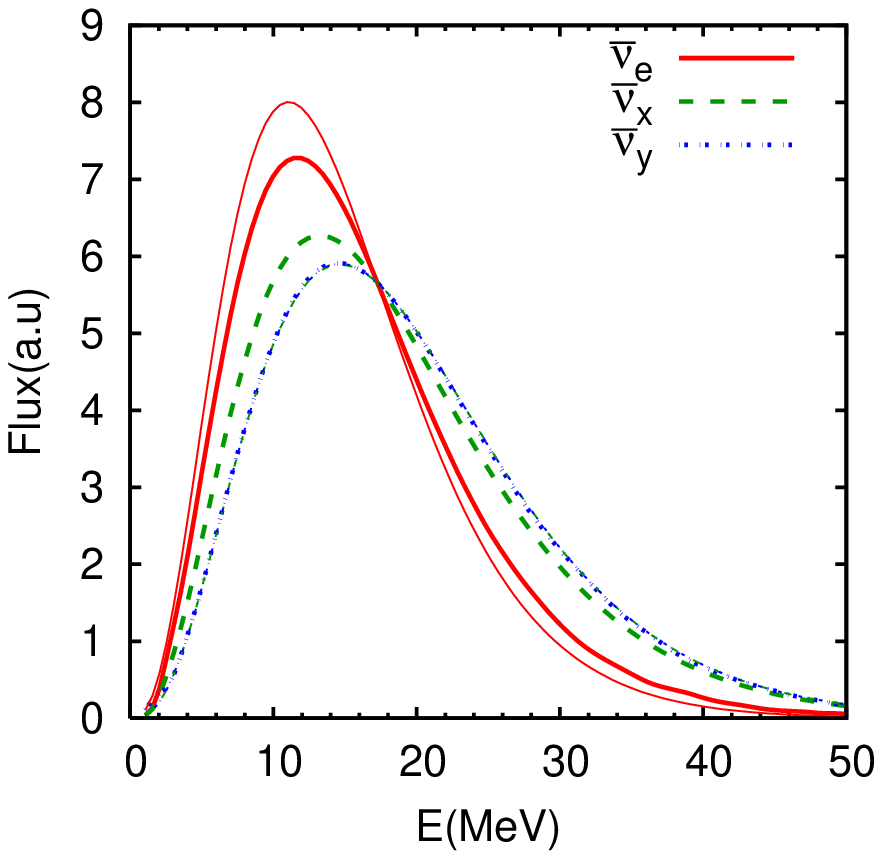,width=7cm}
}
\caption{Neutrino and antineutrino spectra at $r=5000$ km 
for $\theta_{13}=0.1$. The $e,\;x$ and $y$ flavors are shown in red(solid), green(dashes) and blue(dots). The thin lines/dashes/dots are for initial spectra and thick ones for the final spectra. The $\nu_e$ and $\nu_y$ spectra get swapped for  $E \gsim 7$ MeV,
whereas the lower energy $\nu_e$ spectrum partially mixes with
$\nu_x$. In the antineutrino sector, the $\nuebar$ and $\nubar_x$ spectra are partially mixed, while the $\bar{\nu}_y$ spectrum remains unaffected.}
\label{spectra-larget13}
\end{figure}

The neutrino and antineutrino spectra at $r=5000$ km
are shown in Fig.~\ref{spectra-larget13}.
We see that the neutrino spectra have the same
characteristics as for small $\theta_{13}$.
In the antineutrino sector, complete 
$\nuebar$--$\nubar_y$ spectral split and 
the reconversion at the $H$ resonance cancel each other,
whereas the large value of $\theta_{12}$ partially
mixes the $\nuebar$--$\nubar_x$ spectra.

The value of $\theta_{13}$ thus affects the $\nuebar$ spectra
substantially. At larger $\theta_{13}$ values, where the $H$
resonance is more adiabatic, the $\nuebar$ spectrum is softer.
The $\nubar_x$ spectrum is also affected at large $\theta_{13}$,
as opposed to the small $\theta_{13}$ case.

\subsubsection{Summarized results}

It is thus clear that the neutrino fluxes that reach Earth from a SN, are very different from the primary fluxes. In particular for inverted hierarchy, we learn that the $\nue$ and $\nuy$ spectra are exchanged above a certain split-energy $E_c$ due to collective effects. For antineutrinos the swap occurs over the complete spectrum. In the normal hierarchy, collective effects do not have a significant effect. The MSW conversions cause further flavor conversions, and while the conversion probabilities have not changed from the traditional expectation, the primary fluxes entering the resonances are now vastly different. This leads to different flavor composition of the fluxes of neutrinos and antineutrinos arriving on Earth, than was traditionally expected. These fluxes can be calculated using our understanding of collective effects and the level-crossing diagrams. At the detectors on Earth one is typically sensitive to the $\nue$ and/or $\nuebar$ flux, and so we summarize the expectations for these fluxes in Table~\ref{summ-table}.
The expressions in the table are able to describe all the features of
$\nu_e$ and $\bar{\nu}_e$ spectra in Figs.~\ref{spectra-smallt13}
and \ref{spectra-larget13}.

\begin{table}[ht]
\begin{tabular}{lcl}
\hline
Normal hierarchy & $\phantom{space}$ & Inverted hierarchy \\ 
\hline
$F_{\nue}^{obs}= s_{12}^2\left(P_H F_{\nue}+(1-P_H) F_{\nuy}\right) + c_{12}^2 F_{\nux}$  & &
$F^{obs}_{\nue}= \Bigg\{ \begin{array}{lc}
s_{12}^2 F_{\nue}+c_{12}^2F_{\nux} & (E<E_c) \\
s_{12}^2 F_{\nuy}+c_{12}^2F_{\nux} & (E>E_c)\\ 
\end{array} $ \\
\vspace{0.1cm}
$F^{obs}_{\nuebar}= c_{12}^2F_{\nuebar} + s_{12}^2F_{\nuxbar}$ &  & 
$F^{obs}_{\nuebar}= s_{12}^2F_{\nuxbar} + c_{12}^2 \left((1-P_H)F_{\nuebar} + P_H F_{\nuybar}\right)$\\
\hline
\end{tabular}
\caption{Neutrino and antineutrino fluxes arriving on Earth from a SN.}
\label{summ-table}
\end{table}

We have taken the $L$ resonance to be adiabatic. In the case of
multiple $H$ resonances, as may occur during the shock wave
propagation or turbulence, $P_H$ may be taken to be the effective
jump probability (it may have a nontrivial dependence on 
energy and time).
Note that Earth matter effects are present only when 
$F_{\nue/\nuebar}^{obs}$ is a nontrivial combination of 
$F_{\nue/\nuebar}$ and $F_{\nux,\nuy/\nuxbar,\nuybar}$.

\section{Conclusions}                                   
\label{concl}

We have developed a formalism to analyze neutrino flavor
conversion effects in the full three-flavor framework.
It employs the Bloch vector representation for
$3 \times 3$ density matrices, and naturally generalizes
the spin-precession analogy to three flavors.
In particular, it is capable of describing three-flavor
collective neutrino conversion effects inside a core collapse
supernova, like synchronized oscillations, bipolar oscillations 
and spectral split, which have till now been analytically
studied mostly in the two-flavor limit.

We explicitly extend the earlier two-flavor analysis of 
neutrino flavor conversions inside the SN, which includes
neutrino-neutrino interactions, to three flavors,
where we neglect the CP violation in the neutrino sector.
We use the modified flavor basis
$(\nu_e,\nu_x,\nu_y)$, which is rotated from the flavor
basis $(\nu_e, \nu_\mu, \nu_\tau)$ so as to get rid of the
mixing angle $\theta_{23}$.
We also work in the steady state approximation so that
there is no explicit time dependence in the density matrix,
assume spherical symmetry and half-isotropic neutrino
source, and employ the single-angle approximation that
has been shown to be valid in the two-flavor case.
This leads to the equations of a gyroscope in eight
dimensions, similar to the three dimensional gyrosope
equations in the two-flavor case.

In the three-flavor formalism, the density matrix is
represented by an eight-dimensional Bloch vector ${\rm \bf P}$. 
However, the flavor content is determined only by the
two components ${\rm P}_3$ and ${\rm P}_8$ of 
${\rm \bf P}$ after evolution. 
Motivated by this observation, we propose the 
``${\bf e}_3$--${\bf e}_8$'' triangle diagram to represent
the flavor content of any neutrino state by the
projection of ${\rm \bf P}$ on the ${\bf e}_3$--${\bf e}_8$
plane, which we have termed ${\textsf P}$.
This not only allows us to visualize the three-flavor
transformations in a convenient way, but also
allows us to quantify the extent of three-flavor effects
over and above the two-flavor results.

A ``heavy-light'' factorization holds in the
three-flavor treatment for certain initial conditions,
so that the three-flavor results may be understood as 
the two-flavor results with $\msqa$ modified
with terms that depend on $\Delta m^2_\odot$. 
Indeed, in certain situations, the three-flavor neutrino 
conversions may be factorized into three two-flavor
oscillations with hierarchical frequencies. 
In such cases, the three-flavor conversion probabilities
may be constructed from two-flavor results by considering
the modulation of higher frequency modes by lower
frequency modes.

We have compared our analytic results with the numerical ones
for simple cases of an initial pure $\nu_e$ state, 
constant matter densities and no collective
effects, as well as for synchronized oscillations, and 
have found a good agreement even when we ignore the modulation 
due to the lowest frequency.
The additional effect of the third neutrino in these cases
is limited to the excursions of the orbit of ${\textsf P}$
towards $\nu_x$, and has already been well studied (though
without the Bloch vector treatment).
In the absence of collective effects, though the evolution 
of all energies is different, the orbit of ${\textsf P}$ can be 
seen to be an energy-independent quantity.
In the synchronized case, neutrinos of all energies are seen to
oscillate with a common frequency, and even undergo MSW
resonances at the same matter density and with the same 
adiabaticity.

In the case of bipolar oscillations, the addition of the
third neutrino changes the situation significantly.
The analytical results are not so easy to obtain, however the
numerical results for an inverted hierarchy show a ``petal''
pattern in the ${\bf e}_3$--${\bf e}_8$ triangle, 
which can be explained by the combination of $\nu_e
\leftrightarrow \nu_y$ bipolar oscillations and
$\nu_e \leftrightarrow \nu_x$ sinusoidal oscillations. 
The value of the MSW potential also plays an important role
in determining the extent of the effect of the third flavor.
This needs to be explored in more detail.

The spectral split occurs in neutrinos in the inverted hierarchy
when one starts with $\nu_e$, owing to the unstable position of
the eight-dimensional gyroscope in this case.
The $\nu_e$ above a certain energy, and almost all $\nuebar$,
completely convert to $\nu_y$ and $\nubar_y$ respectively.
There are no additional split effects from the introduction of
the third flavor since the hierarchy in the solar sector is normal. 
This, however, could change if neutrinos are not in a pure $\nu_e$
state as they enter the bipolar region.

We have simulated the neutrino flavor conversions numerically by
taking a realistic density profile for the SN,
and have shown the flavor conversions for inverted
hierarchy and two $\theta_{13}$ values in the
cooling phase.
In such a scenario, it is easily possible to identify
regions where different collective as well as MSW
effects dominate.
We are able to predict the regions in which these effects
take place, and our three-flavor formalism can explain
the features of flavor conversions therein.
We also point out an interplay between the collective and
MSW effects. For example,
the $H$ resonance cancels the effect of the spectral split
for antineutrinos, whereas
the spectral split makes the $L$ resonance irrelevant
for neutrinos above the split energy. 
If the hierarchy were normal, the collective effects would be
effectively absent in the cooling phase.

In conclusion, a complete understanding of the neutrino flavor 
conversions inside a SN requires a three-flavor treatment.
In this paper, we have developed a formalism to handle this
analytically, provided a method to estimate the three-flavor
probabilities using the two-flavor results in certain situations, 
and have pointed out an interplay between collective and
MSW effects, which can be easily understood with the formalism.


\section*{Acknowledgements}

We would like to thank K. Damle, S. Gupta, R. Loganayagam,
A. Mirizzi, G. G. Raffelt 
for useful discussions and insightful comments,
and V. Tripathi for factors of three.
This work is partly supported through the
Partner Group program between the Max Planck Institute
for Physics and Tata Institute of Fundamental Research.


\appendix 

\section{Notation}

This paper uses 3-vectors, $3 \times 3$ matrices, 8-vectors,
their components and their projections in two-dimensional planes.
We have tried to be consistent in the use of fonts for all these
objects. 
The convention followed for indices is:
\begin{itemize}
\item $i,j,k \in \{1,2,3 \}$ index mass eigenstates
\item $\alpha,\beta \in \{e,x,y \}$ index modified flavor eigenstates
\item $a,b,c \in \{1,2,3,4,5,6,7,8\}$ are SU(3) indices.
\end{itemize}
The convention followed for symbols is:
\begin{itemize}
\item Three-dimensional vectors are denoted by smallcase boldfaced
letters, e.g. ${\bf p,q,r,v}$. 
\item $3 \times 3$ matrices are denoted in the regular math mode,
e.g. $\rho, H, I, M, O, R_{ij}, U, V, \Lambda_a$. Note however the exceptions, energy $E$ and the Fermi constant $G_F$, that are also respresented in this font. 
\item Sets of $3 \times 3$ matrices are denoted by blackboard-bold
font, e.g. ${\mathbbm K}, {\mathbbm Q}$.
\item Eight-dimensional vectors are written in bold capital letters,
e.g. $\Bvec, \Dvec, \Pvec, \Svec, \Xvec$.
Exceptions are ${\bf e}_a$, which are vectors along the coordinate axes,
and therefore conventionally written in smallcase.
\item The components of an eight-vector $\Xvec$ are written as ${\rm X}_a$. 
\item ${\bm\Lambda}$ is a eight-dimensional vector whose components are
$3\times 3$ matrices $\Lambda_a$.
\item Projections of eight-vectors on the ${\bf e}_3$--${\bf e}_8$ 
plane, e.g. ${\textsf P}$, as well as the rotation and evolution
matrices on that plane, e.g. ${\textsf R}, {\textsf S}$ are written in
sans-serif font.
\end{itemize}


\end{document}